\documentclass[aps,pre,showpacs,superscriptaddress,groupedaddress,twocolumn]{revtex4}
\usepackage{graphicx}  
\usepackage{bm}        
\usepackage{amssymb}   
\usepackage{amsmath}
\usepackage{enumerate}
\usepackage{color}
\usepackage{pgfplots}
\usepackage{tikz}
\usetikzlibrary{calc,decorations.markings}

\begin{document}

\title{Continuous information flow fluctuations}
\author{ M.L. Rosinberg}
\affiliation{Laboratoire de Physique Th\'eorique de la Mati\`ere Condens\'ee, Universit\'e Pierre et Marie Curie,CNRS UMR 7600,\\ 4 place Jussieu, 75252 Paris Cedex 05, France}
\email{mlr@lptmc.jussieu.fr}
\author{Jordan M. Horowitz}
\affiliation{Physics of Living Systems Group, Department of Physics, Massachusetts Institute of Technology - 400 Technology Square, Cambridge, MA 02139}

\begin{abstract}
Information plays a pivotal role in the thermodynamics of nonequilibrium processes with feedback.  
However, much remains to be learned about the nature of information fluctuations in small scale devices and their relation with fluctuations in other thermodynamics quantities, like heat and work. Here we derive a series of fluctuation theorems for information flow and partial entropy production in a Brownian particle model of feedback cooling and  extend them to arbitrary driven diffusion processes. We then analyze the long-time behavior of the feedback-cooling model in detail. Our results provide insights into the structure and origin of large deviations of information and thermodynamic quantities in autonomous Maxwell's demons.

\end{abstract} 

\pacs{05.70.Ln,05.40.-a,89.70.-a}

\maketitle

\section{Introduction}

Accounting for information is necessary to rationalize thermodynamics in the presence of feedback~\cite{PHS2015}. In small systems where noise is unavoidable, information not only  bounds the average extracted work, but also fluctuates along individual stochastic trajectories, alongside heat and work. Although this feature has been incorporated in generalized (detailed and integral) fluctuation relations~\cite{SU2010,HV2010,SU2012,ES2012,IS2013,HBS2014,KMSP2014,SS2015}, little is known about the properties of information fluctuations and their correlations with other thermodynamic quantities. Of particular interest is the behavior at long times, an issue recently addressed for a discrete two-state information engine~\cite{MGH2015}.

In this Letter, we analyze the large deviation statistics of information fluctuations for a Brownian particle model, which may be viewed as a dynamic version of a Maxwell's demon~\cite{PHS2015}. This model describes a feedback cooling (or cold damping) experiment~\cite{PZ2012} and has proven to be a rich playground for theoretical exploration~\cite{KQ2007,IS2011,MR2012,MR2013,HS2014}.
We begin by proving a series of  transient integral fluctuation theorems (IFTs).  One of them, applicable to  any coupled Langevin processes experiencing independent noises, is the analog of the IFT for bipartite Markov jump processes derived in \cite{SS2015}.
We thereby confirm that we have identified the correct fluctuating analog of information flow in diffusion processes. With this groundwork, we calculate analytically and numerically the information flow fluctuations  in the steady-state regime and their correlations with heat.
This analysis reveals strong correlations that extend beyond the average behavior into the large, rare fluctuations regime. 
Further insight is then gained by unraveling the atypical trajectories that lead to such rare information or heat fluctuations.

\section{Setup}

Consider a one-dimensional underdamped Brownian particle of mass $m$ immersed in a thermal environment with viscous damping $\gamma$ and temperature $T$.
Feedback cooling is implemented by measuring the particle's velocity $v_t$ using a low-pass filter with cut-off frequency $1/\tau$, and feeding back the measurement outcome $y_t$ with gain $\kappa$ as an additional friction force $f_t=-\kappa y_t$.
The resulting dynamical evolution, including measurement and control, is captured by the coupled Langevin equations~\cite{HS2014} 
\begin{align}
\label{Eqdef} 
m\dot v_t&=-\gamma v_t -\kappa y_t+\xi_t\nonumber\\
\tau \dot y_t&=-(y_t -v_t-\eta_t),
\end{align}
where $\xi_t$ is  Gaussian thermal noise with zero mean and variance $\langle \xi_t\xi_{t'}\rangle =2\gamma T\delta (t-t')$, and $\eta_t$ is Gaussian measurement noise with zero mean and variance $\langle \eta_t\eta_{t'}\rangle =\Delta\delta (t-t')$.
Here and throughout  Boltzmann's constant   is set to unity.  An equivalent implementation of (1) in an electric circuit is discussed in \cite{SDNM2014} {(see also \cite{MR2013} for a more general version of the model with a  harmonic potential trapping the Brownian particle)}.

The feedback's purpose is to maintain the system in a nonequilibrium steady state (NESS), where the average kinetic temperature $T_{\rm kin}\equiv m\langle v^2\rangle_{\rm st}$ is  smaller than $T$ ({the subscript ``st" indicates that the average is taken in the stationary state of the joint process)}.
Consequently, the feedback controller must be extracting energy from a single heat reservoir and converting it into work, in apparent violation of the second law. 
However,  the controller acts as a Maxwell's demon that autonomously gathers information in order to implement the feedback.
This information saves the second law by providing a rigorous bound on the extracted work through any of several second-law-like inequalities, each utilizing a different notion of information~\cite{HS2014}.
We here focus on the ``information flow"~\cite{AJM2009,HE2014} (or learning rate~\cite{BHS2014,HBS2016}) and its fluctuations.

\section{Information flow on the trajectory level}

Stochastic information flow is a trajectory-level quantity that captures the dynamic variation
 of the correlations between $y_t$ and $v_t$.
To define this information flow, we require the time-dependent probability density $p_t(v,y)$, which evolves according to the Fokker-Planck equation~\cite{HS2014}
\begin{equation}\label{eq:FP}
d_t p_t(v,y)=-\partial_v J^v_t(v,y)-\partial_y J^y_t(v,y),
\end{equation}
with probability currents
\begin{align}
J^v_t(v,y) &=- \frac{1}{m}(\gamma v+\kappa y) p_t(v,y)-\frac{\gamma T}{m^2}\partial_vp_t(v,y)\\
J^y_t(v,y) &=- \frac{1}{\tau}(y-v) p_t(v,y)-\frac{\Delta}{2\tau^2}\partial_vp_t(v,y).
\end{align}
Consequently, the time derivative of the marginals is {$d_t p_t(v)=-\partial_v J^v_t(v)$ with  $J^v_t(v)=\int dy\: J_t^v(v,y)$}, and similarly for $p_t(y)$.

{We are here interested in the information flow due to the $v$-fluctuations, whose average enters the generalized second-law inequality\cite{HS2014}.} This quantity is defined as the partial rate of change of the stochastic mutual information 
\begin{align}
I_t(v_t:y_t)=\ln \frac{p_t(v_t,y_t)}{p_t(v_t)p_t(y_t)}\ .
\end{align}
Namely, we split the total time variation of $I_t$  as 
\begin{align} 
\label{EqdtIt}
\frac{dI_t}{dt}=-(\imath_t^v+\imath_t^y)\ ,
\end{align}
into a piece arising due to $v$-fluctuations  
{{\begin{align} 
\label{Eqitv}
\imath_t^v(v_t,y_t)&\equiv \frac{1}{p_t(v_t,y_t)}\partial_v J^v_t(v,y)\vert_{v_t,y_t}-\dot v_t \partial_v \ln p_t(v,y)\vert_{v_t,y_t}\nonumber\\
&-\frac{1}{p_t(v_t)}\partial_v J^v_t(v)\vert_{v_t}+\dot v_t \partial_v \ln p_t(v)\vert_{v_t}
\end{align}}
and a similar piece due to $y$-fluctuations, $\imath_t^y(v_t,y_t)$. {In these expressions, the probabilities and currents  obtained by solving (\ref{eq:FP}), are evaluated along the stochastic trajectories ${\bf v}_0^t\equiv \{v_{t'}\}_{0\le t'\le t}$ and ${\bf y}_0^t\equiv \{y_{t'}\}_{0\le t'\le t}$ generated by (\ref{Eqdef}).}

Like the stochastic Shannon entropy~\cite{S2005}, $\imath_t^v$ has a mixed character as it depends on both the micro-state $(v_t,y_t)$ and  the whole ensemble of trajectories with initial density $p_0(v,y)$. The ensemble average correctly yields the information flow~\cite{AJM2009,HE2014,HS2014}: $\langle \imath_t^v\rangle= -\int dv dy\: J_t^v(v,y)\partial_v \ln p_t(y\vert v)$. (Note that we choose a minus sign in (\ref{EqdtIt}) so that $\langle \imath_t^v\rangle_{\rm st}>0$, like in \cite{HS2014}; the average information flow has an opposite sign in \cite{AJM2009,HE2014}.) By integrating $\imath_t^v(v_t,y_t)$ over the time interval $[0,t]$, we introduce the trajectory observable, the integrated information current
\begin{align} 
\label{EqDeltaI}
I^v\equiv\int_0^t dt'\:\imath_{t'}^v(v_{t'},y_{t'})= \int_0^t dt'\: \dot s_{t'}^v(v_{t'},y_{t'})+\ln \frac{p_t(v_t)}{p_0(v_0)}\ ,
\end{align}
where ${\dot s}^v_t(v_t,y_t)=\partial_v J^v_t(v_t,y_t)/p_t(v_t,y_t)-\dot v_t \partial_v \ln p_t(v_t,y_t)$ is the time-variation of the stochastic joint entropy $s_t(v_t,y_t)=-\ln p_t(v_t,y_t)$ due to $v$'s dynamics.

\section{Fluctuation theorems}

The introduction of the time-integrated information current $I^v$ allows us to generalize the integral fluctuation theorem (IFT) for the stochastic entropy production in the presence of continuous feedback. 
For an observer unaware of the existence of the controller, the apparent ``total" entropy production in the time interval $[0,t]$ would be
\begin{align}
\label{Eqsigapp} 
\Sigma^v\equiv \Delta s- \frac{{\cal Q}}{T}\ ,
\end{align}
where $\Delta s=-\ln p_t(v_t)+\ln p_0(v_0)$ is the entropy change in the system~\cite{S2005}, and ${\cal Q}=\int_0^t dt'\:(-\gamma v_{t'}+\xi_{t'})\circ v_{t'}$ is the heat received by the particle from the bath~\cite{S1997}, corresponding to an entropy change in the medium $\Sigma^m=-{\cal Q}/T$ (the symbol $\circ$ denotes a Stratonovich integral). However, as a result of the feedback, $\Sigma^v$  is negative on average in the stationary cooling regime, and more generally does not verify a fluctuation theorem, $\langle e^{-\Sigma^v}\rangle\ne 1$. 
Missing in the exponential is the entropy change provided by the feedback mechanism. 
Guided by the case of Markov jump processes, we introduce a new observable, dubbed a ``partial" entropy production~\cite{SS2015},
\begin{align}
\label{EqSigma} 
\Sigma\equiv \Sigma^v+ I^v=\Sigma^m+\int_0^t dt'\:\dot s_{t'}^v(v_{t'},y_{t'})\ ,
\end{align}
and demonstrate that it obeys the generalized IFT
\begin{align}
\label{EqIFT1} 
\left\langle e^{-\Sigma}\right\rangle=1Ê\ ,
\end{align}
{where $\left\langle ...\right\rangle$ denotes an average over all possible paths of duration $t$ with initial state drawn from a distribution $p_0(v_0,y_0)$}. This is a central result of this Letter. By Jensen's inequality, one recovers the second-law-like inequality $\langle \Sigma^v+ I^v\rangle\ge 0$. In particular, the work extracted in the steady state  obeys $\langle W_{\rm ext}\rangle_{\rm st}=-T\langle \Sigma^m\rangle_{\rm st}\le T \langle I^v\rangle_{\rm st}$ {(as $\left\langle \Delta s\right\rangle_{\rm st}=0$ and thus $\left\langle \Sigma^v\right\rangle_{\rm st}=\left\langle\Sigma^m\right\rangle_{\rm st}$)}.

Although $\Sigma$ is not a coarse-grained observable since it is a functional of both trajectories ${\bf v}_0^t$ and ${\bf y}_0^t$, it is nontrivial that it satisfies an IFT.  To prove this result, we will show that $\Sigma$ can be cast as the log-ratio of the probability ${\cal P}[{\bf v}_0^t,{\bf y}_0^t]$ of observing the trajectory $({\bf v}_0^t,{\bf y}_0^t)$ to the probability  ${\cal P}^*[\tilde {\bf v}_0^t, \tilde {\bf y}_0^t]$ of observing the time-reversed trajectory $(\tilde{\bf v}_0^t,\tilde{\bf y}_0^t)=(-{\bf v}_t^0,{\bf y}_t^0)$  in a suitable defined {\it modified} dynamics:
\begin{align}
\label{EqSigma1}
\Sigma=\ln \frac{{\cal P}[{\bf v}_0^t,{\bf y}_0^t]}{{\cal P}^*[\tilde {\bf v}_0^t, \tilde {\bf y}_0^t]}.
\end{align}
The modified dynamics only alters the measurement process and is generated by
\begin{align}
\label{Eqstar} 
\tau \dot y_t=y_t +v_t+\frac{\Delta}{\tau}\partial_y\ln p_t(-v_t,y_t)+\eta_t.
\end{align}
This dynamics is intimately related to the {\it auxiliary} or {\it driven} process~\cite{JS2010,GL2010,CT2013} that generates  the constrained path ensemble ${\cal P}[{\bf v}_0^t,{\bf y}_0^t\vert \Sigma=\sigma t]$ at long times (see  (\ref{Eqeff}) below).

{The proof of (\ref{EqSigma1}) proceeds by first writing out (\ref{EqSigma1}) as
\begin{align}
\label{EqSigma2}
\Sigma=\ln \frac{{\cal P}[ {\bf  v}_0^t\vert  {\bf y}_0^t,v_0]{\cal P}[ {\bf  y}_0^t\vert  {\bf v}_0^t,v_0]p_0(v_0,y_0)}{{\cal P}[\tilde {\bf v}_0^t\vert \tilde {\bf y}_0^t,{\tilde v}_0]{\cal P}^*[\tilde {\bf y}_0^t\vert \tilde {\bf v}_0^t,{\tilde y}_0]p_t(v_t,y_t)}\ ,
\end{align}
where the  path probabilities ${\cal P}[{\bf v}_0^t\vert  {\bf y}_0^t, v_0]$, ${\cal P}[{\bf y}_0^t\vert  {\bf v}_0^t, y_0]$, etc. are expressed in terms of Onsager-Machlup action functionals~\cite{note2}. For instance, 
\begin{align}
\label{EqOMy}
{\cal P}[{\bf y}_0^t\vert {\bf  v}_0^t,y_0]&\propto e^{-\frac{t}{2\tau}} e^{-\frac{1}{2\Delta}\int_0^t dt'\:[\tau \dot y_{t'}+y_{t'}-v_{t'}]^2}\ ,
\end{align}
using the Stratonovich discretization. (As stressed in \cite{HS2014}, the path functionals in (\ref{EqSigma2}) are not true conditional probabilities because $v_t$ and $y_t$ influence each other.)  The conclusion follows  by using the local detailed balance relation $\ln \big({\cal P}[{\bf  v}_0^t\vert {\bf y}_0^t,v_0]/{\cal P}[\tilde  {\bf v}_0^t\vert \tilde  {\bf y}_0^t,{\tilde v}_0]\big)=\Sigma^m$ and noting that the dynamics generated by \eqref{Eqstar} is absolutely continuous with respect to \eqref{Eqdef}
 \begin{equation}
{\cal P}^*[\tilde {\bf y}_0^t\vert \tilde {\bf v}_0^t,{\tilde y}_0]={\cal P}[ {\bf  y}_0^t\vert  {\bf v}_0^t,v_0]e^{\int_0^tdt^\prime\, \left[\frac{d}{dt}s_{t^\prime}(v_{t^\prime},y_{t^\prime})-{\dot s}^v_{t^\prime}(v_{t^\prime},y_{t^\prime})\right]}.
 \end{equation}
Details can be found in the Supplemental Material (SM).}
As $\Sigma$ is a log-ratio of path probabilities, one  readily obtains the IFT (\ref{EqIFT1}).

{Note that the dynamics of  $v_t$ does not play any role in this calculation and that the force acting on $y_t$ could be  arbitrary. In other words, the IFT holds for any coupled Langevin processes involving independent noises~\cite{JH2015} (as it holds for any bipartite Markov jump processes~\cite{SS2015}). Consider for instance the two coupled overdamped Langevin equations
\begin{align} 
\dot x_t&=\mu_xF_1(x_t,y_t,t)+\xi_t^x\nonumber\\
\dot y_t&=\mu_y F_2(x_t,y_t,t)+\xi_t^y\ ,
\end{align}
 where $\langle \xi_t^i\xi_{t'}^j\rangle= 2D_i\delta_{ij}\delta (t-t')$ and $D_i=T_i\mu_i$. Then, replacing $F_2(x_t,y_t,t)$ by
\begin{align} 
F_2^*(x,y,t)=-F_2(x_t,y_t,t)+2T_y\partial_y\ln p_t(x_t,y_t)\ ,
\end{align}
one can derive an IFT like \eqref{EqIFT1} by a similar argument (see SM).  
Of course, a similar IFT holds for the partial entropy production of the $y_t$ degree of freedom.}

Two other fluctuation theorems are obtained in a similar manner by comparing the original feedback process to other modified dynamics. First, by flipping the sign of the viscous damping $\gamma \to -\gamma$ in the first Langevin equation, 
\begin{align}
\label{Eqstarhat} 
m\dot v_t&=\gamma v_t -\kappa y_t+\xi_t\ ,
\end{align}
{we obtain an IFT for the dissipated heat  (or medium entropy production $\Sigma^m$)
\begin{align}
\label{EqIFT3} 
\left\langle e^{-\Sigma^m}\right\rangle=e^{\frac{\gamma}{m}t}Ê\ .
\end{align}
This relation, originally derived in \cite{RTM2016}, holds for any underdamped Langevin dynamics, provided the damping is linear. Finally, by  combining the two dynamics  (\ref{Eqstar}) and (\ref{Eqstarhat}), we find an IFT that includes information (see SM)
\begin{align}
\label{EqIFT2} 
\left\langle\frac{p_t(v_t)}{p_0(v_0)}e^{-I^v}\right\rangle=e^{\frac{\gamma}{m}t}Ê\ .
\end{align}}

\section{Fluctuations in the NESS}

We now turn to calculating the stationary-state fluctuations of the integrated information current $I^v$ and the medium entropy flow (or heat) $\Sigma^m$ in our linear feedback cooling model \eqref{Eqdef}.
To simplify the notation, we drop the subscript ``st". 

Thanks to the linearity of (\ref{Eqdef}), the stationary distribution of the joint system is Gaussian 
\begin{align} 
\label{Eqpst}
p(v,y)=\frac{1}{\sqrt{(2\pi^2)\vert {\bf C}\vert}}e^{-\frac{1}{2}(v,y).{\bf C}^{-1}.(v,y)^T}\ ,
\end{align} 
where the entries of the covariance matrix  {${\bf C}$} ($c_{11}= \langle v^2\rangle,\, c_{12}= \langle vy\rangle,\, c_{22}=\langle y^2\rangle$) are given  in Appendix D of \cite{HS2014}. 
From (\ref{Eqitv}) and (\ref{EqDeltaI}), we then obtain the explicit expression for $I^v$:
\begin{align} 
\label{Eqiv}
I^v&=-A_1t+ \frac{1}{2}(\alpha_{11}-\frac{1}{\sigma_{11}})(v_t^2-v_0^2)\nonumber\\
&+\int_0^t dt'\:(\alpha_{11}A_1v_{t'}^2+A_2 y_{t'}^2+A_3v_{t'}y_{t'}+\alpha_{12}\dot v_{t'}y_{t'})\ ,
\end{align}
where $\alpha_{11}=c_{22}/\vert {\bf C}\vert, \alpha_{12}=-c_{12}/\vert {\bf C}\vert$, and  
\begin{align} 
\label{EqcoeffA}
A_1&=\frac{\gamma}{m}(1-T\frac{\alpha_{11}}{m})\nonumber, \quad A_2=\frac{\alpha_{12}}{m}(\kappa-\frac{\gamma T}{m}\alpha_{12})\nonumber\\
A_3&=\frac{a}{m}\alpha_{11}+\frac{\gamma}{m}\alpha_{12}(1-2T\frac{\alpha_{11}}{m})\ .
\end{align}

\subsection{Large deviation analysis}

As $t\to \infty$, the stationary probability distribution of a time-integrated observable ${\cal A}$ -- such as $I^v$ or $\Sigma^m$-- is said to satisfy a large deviation principle if it  takes the scaling form $P({\cal A}=a t)Ê\sim e^{-tE(a)}$, where $E(a)$ is the large deviation rate function (LDF)~\cite{T2009}. As usual, it is convenient to  introduce the associated moment generating function $Z_{a}(\lambda,t)=\langle e^{-\lambda {\cal A}}\rangle$, which behaves asymptotically as 
\begin{align} 
\label{EqZi}
Z_a(\lambda,t)\sim g_a(\lambda) e^{t\mu_a(\lambda)}\ ,
\end{align}
where $\mu_a(\lambda) \equiv \lim_{t \to \infty} (1/t) \ln Z_a(\lambda,t)$ is the scaled cumulant generating function (SCGF) and $g_a(\lambda)$ is a sub-leading factor. The LDF is  normally obtained via the Legendre transform $E(a)=-[\mu_a(\lambda^*(a))+\lambda^*(a) a]$, with the saddle point $\lambda^*$ determined by $\mu_a'(\lambda^*(a))=-a$~\cite{T2009}. This relation, however, breaks down if  $g_a(\lambda)$ has a singularity in the region of the saddle-point integration, due to rare but large fluctuations of a boundary temporal term ({\em e.g.}\ the second term in the first line of (\ref{Eqiv})). The leading contribution to the LDF then comes from the singularity, which induces an exponential tail in the pdf.  As stressed in \cite{RTM2016}, this may even {make the SCGF discontinuous at $\lambda=1$} when the modified process, such as the one governed by (\ref{Eqstar}), has no stationary density. 
 
We calculate  the SCGFs {for the medium entropy production rate $\sigma^m=\Sigma^m/t$ and the information flow $\imath^v=I^v/t$} by direct integration of the path probability. The calculation is made tractable by imposing periodic boundary conditions on the trajectories, which allows us to expand $v_t$ and $y_t$ in a discrete Fourier series (see {\em e.g.}\ \cite{ZBCK2005,KSD2011}). The results are conveniently expressed in terms of the response function in the frequency domain 
\begin{align} 
\chi(\omega)=\frac{1-i\omega \tau}{\kappa+\gamma-i(m+\gamma \tau)\omega -m\omega^2\tau}\ ,
\end{align}
and two auxiliary functions
\begin{subequations}
\label{F:subeqns}
\begin{align} 
&F_{\sigma^m,\lambda}(\omega)=\frac{2\gamma \Delta \kappa^2}{T}\frac{\vert \chi(\omega)\vert^2}{1+\omega^2\tau^2}\Big(1-\frac{2T}{\Delta \kappa}-\lambda\Big)\lambda \label{F:subeq1}\\
&F_{\imath^v,\lambda}(\omega)=2\gamma T\Delta \alpha_{12}^2\frac{\vert \chi(\omega)\vert^2}{1+\omega^2\tau^2} \Big[\Big(\frac{\kappa}{m}\frac{\alpha_{11}}{ \alpha_{12}}-\frac{\gamma}{m}\Big)^2+ \omega^2\Big]\lambda^2 \label{F:subeq2}\ ,
\end{align}
\end{subequations} 
as (see SM)
\begin{subequations}
\label{mu:subeqns}
\begin{align} 
&\mu_{\sigma^m}(\lambda)=-\int_{0}^{\infty}\frac{d\omega}{2\pi} \ln [1-F_{\sigma^m,\lambda}(\omega)]\label{mu:subeq1}\\
&\mu_{\imath^v}(\lambda)=A_1\lambda-\int_{0}^{\infty}\frac{d\omega}{2\pi} \ln [1-F_{\imath^v,\lambda}(\omega)]\label{mu:subeq2}\ .
\end{align}
\end{subequations} 
 We highlight two important features of (\ref{F:subeqns})-(\ref{mu:subeqns}). First, $\mu_{\sigma^m}(\lambda)$ has the  symmetry $\mu_{\sigma^m}(\lambda)=\mu_{\sigma^m}(1-\frac{2T}{\Delta \kappa}-\lambda)$, which implies that the pdf $P(\Sigma^m)$ satisfies the steady-state fluctuation theorem 
 \begin{align} 
\label{EqFT}
\lim_{t\to \infty}\frac{1}{t}\ln \frac{P(\Sigma^m=\sigma^mt)}{P(\Sigma^m=-\sigma^mt)}=\left(1-\frac{2T}{\Delta \kappa}\right)\sigma^m\ ,
\end{align}
at least in a limited range of $\sigma^m$ around $0$ (see Fig. \ref{FigLDF} below).
Second, $F_{\imath^v,\lambda}(\omega)$ is an even function of $\lambda$. Therefore, the LDF $E(\imath^v)$ is symmetric around the expectation value  $\langle \imath^v\rangle=-\mu'_{\imath^v}(0)=-A_1$. 
We will elaborate on this point below.
(On the other hand, the SCGF of $\Sigma$, given by a similar but more complicated expression - see SM -, does not display any symmetry.)

To study the correlations between  $\Sigma^m$ and $I^v$  in the long-time limit, we also compute by the same method the SCGF $\mu_{\sigma^m,\imath^v}(\lambda_1,\lambda_2)=\lim_{t \to \infty}(1/t)\ln \langle e^{-\lambda_1 \Sigma^m-\lambda_2 I^v} \rangle$ and the corresponding   joint LDF $E(\sigma^m,\imath^v)$ (see SM). As an added benefit, this  allows us to investigate the fluctuations of the ratio $\epsilon=-\Sigma^m/I^v$ that characterizes  the efficiency of the information-to-work conversion along the trajectories.
The most probable value of $\epsilon$ is the ``macroscopic" efficiency $\bar{\epsilon}=-\langle \sigma^m\rangle/\langle \imath^v\rangle$~\cite{CF2009,ES2012,BAS2012}. 

\subsection{Numerical study}

To further explore the fluctuations of trajectory observables, we now present some numerical results. Hereafter the model is described by three dimensionless parameters: the feedback gain $g=\kappa/\gamma$,  the signal-to-noise ratio $\mbox{SNR}=2T/(\gamma \Delta)$, and the ratio $\tau/\tau^v$ where $\tau^v=m/\gamma$ is the velocity relaxation time. Specifically, we set $\tau/\tau^v=0.01$ and $\mbox{SNR}=40$  and vary the feedback gain $g$, as in experiments~\cite{PZ2012}. By choosing a moderate noise level, we make the control more sensitive to information-flow fluctuations~\cite{B2015}.
\begin{figure}[hbt]
\begin{center}
\includegraphics[trim={0 1.25cm 0 1.75cm},clip,width=7cm]{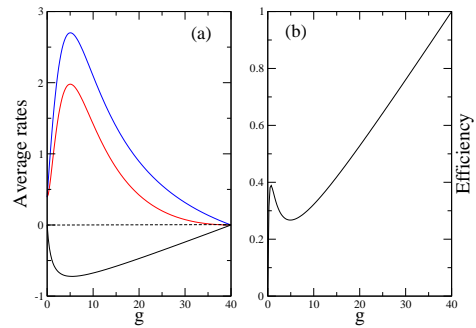}
\caption{ (Color on line) (a) The average rates $\langle\sigma^m\rangle$ (black line), $\langle\sigma\rangle$ (red line),  and  $\langle \imath^v\rangle$ (blue line) as a function of the feedback gain $g$. One has $\langle \sigma\rangle=\langle\sigma^m\rangle+\langle \imath^v\rangle\ge 0$. (b) The most probable efficiency $\bar\epsilon=-\langle\sigma^m\rangle/\langle \imath^v\rangle$ as a function of $g$.}
\label{Figrates}
\end{center}
\end{figure}

Let us first recall that  cooling [$T_{\rm kin}<T$ and thus $\langle \sigma^m\rangle=(1/\tau^v)(T_{\rm kin}-T)<0$]  requires  $g/\mbox{SNR}<1$, independent of the value of $\tau$ (cf. eq. (20) in \cite{HS2014} with the noise variance $\sigma^2$ replaced by $\Delta$).  This is illustrated in Fig. \ref{Figrates}(a) where we plot  the average rates as a function of $g$. The most salient features are the extrema in $\langle \sigma^m\rangle$ and $\langle \imath^v\rangle$. The minimum in $\langle \sigma^m\rangle$ occurs at {{$g=g_{\rm opt}=\sqrt{1+\mbox{SNR}}-1$}{~\cite{HS2014}}.  Above $g_{\rm opt}$, too much measurement noise is fed back to the system, causing $T_{\rm kin}$ to increase with $g$, a well-known experimental fact~\cite{PZ2012}. Eventually, for $g/\mbox{SNR}>1$, the system is heated instead of cooled. 
On the other hand, {for $\tau\ne 0$}, the maximum in the  information flow occurs at {$g_{\rm KB}=g_{\rm opt} [1-(\tau/\tau^v)\sqrt{1+\mbox{SNR}}]<g_{\rm opt}$}. The demon then realizes a {\it Kalman-Bucy} filter{~\cite{AM2008}} (hence the notation $g_{\rm KB}$).
In this limit, $\hat v_t\equiv  (\tau/\tau^v) g_{\rm opt}y_t$ represents  the best estimate of $v_t$ in terms of the mean-squared error ${\cal E}_t=\langle (v_t-\hat v_t)^2\rangle$, given all past measurements~\cite{HS2014,SDNM2014} (recall that (\ref{Eqdef}) describes a non-Markovian control protocol~\cite{SU2012}). Interestingly, {as shown in \cite{HS2014},} $\langle \imath^v\rangle$ is then equal to  the transfer entropy rate $g_{\rm opt}/2$. We see in Fig. \ref{Figrates}(b) that these extrema in $\langle \sigma^m\rangle$ and $\langle \imath^v\rangle$ induce a local minimum in the information efficiency $\bar{\epsilon}$. This minimum would exactly occur at $g_{\rm opt}$ if $\tau$ were zero (note  that $\bar \epsilon=1$ for $g/\mbox{SNR}=1$, when the demon does not extract any work.). 
\begin{figure}[hbt]
\begin{center}
\includegraphics[trim={0 1.25cm 0 1.75cm},clip,width=7cm]{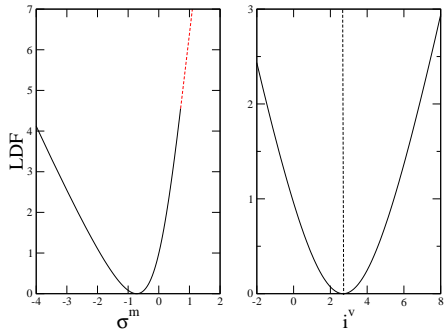}
\caption{ (Color on line) Large deviation functions $E(\sigma^m)$ (left) and $E(\imath^v)$ (right) at maximum power ($g=g_{\rm opt}$). The FT symmetry $E(\sigma^m)-E(-\sigma^m)=(\frac{2T}{\Delta \kappa}-1)\sigma^m$ is obeyed for $ \vert \sigma^m\vert< 0.712$ whereas $E(\sigma^m)$ is linear for $\sigma^m >0.712$  (dashed red line). $E(\imath^v)$ is symmetric around its minimum.}
\label{FigLDF}
\end{center}
\end{figure}

We now fix the gain at its optimal value $g_{\rm opt}$ and investigate the fluctuations. We stress that $g_{\rm opt}$ is  close to  $g_{\rm KB}$ with our choice of the parameters, so that $\langle \imath^v\rangle$ is almost maximal (and the so-called sensory capacity~\cite{HBS2016} is close to $1$). The large deviation functions  $E(\sigma^m)$ and $E(\imath^v)$  are plotted in Fig. \ref{FigLDF}. Notice that $E(\sigma^m)$ has a linear branch for $\sigma^m>0.712$ due to the presence of a pole in the pre-exponential factor. As a result, the steady-state FT (\ref{EqFT}) does not hold for large values of $\sigma^m$ (this, however, corresponds to extremely rare events and depends on the choice of the model parameters). More intriguing is the symmetry exhibited by $E(\imath^v)$ around its minimum, which we  already pointed out.  This implies that positive and negative fluctuations of $I^v$ around the expectation value are equiprobable.
While we have no complete analytic proof {nor heuristic  argument}, we believe this symmetry is a general property of our model even for finite-time fluctuations.
 This is indeed suggested by numerical simulations as well as by a small-$t$ expansion of the modified generating function $Z_{\imath^v}(\lambda,t)e^{\lambda \langle \imath^v\rangle t}=\langle e^{-\lambda [I^v-\langle I^v\rangle]}\rangle$  displaying only even powers of $\lambda$ (see SM). 
 It should be added that perturbative calculations show that the symmetry is lost when there are nonlinearities in the feedback control~\cite{note1}.

{
\begin{figure}
\centering
\includegraphics[trim={0 1.25cm 0 1.75cm},clip,width=7cm]{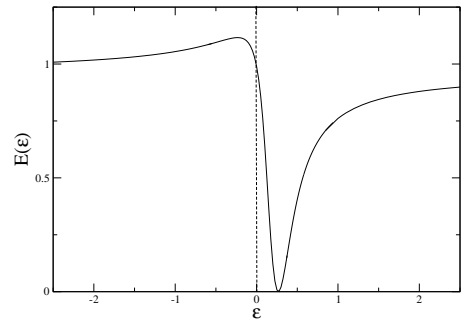}
\caption{Large deviation function of the efficiency $E(\varepsilon)$ for  $g=g_{\rm opt}$.  The most and least probable values are $\varepsilon_{\rm most}=\bar \epsilon\approx0.268$ and $\varepsilon_{\rm least}\approx -0.225$.}
\label{FigEepsilon}
\end{figure}
The LDF for the efficiency $\epsilon$ is obtained from the joint SCGF $\mu_{\sigma^m,\imath^v}(\lambda_1,\lambda_2)$ as $E(\epsilon)=-\inf_{\lambda_1}\mu_{\sigma^m,\imath^v}(\lambda_1,\epsilon \lambda_1)$ and  is plotted in Fig. \ref{FigEepsilon}.
 Like in the case of stochastic heat engines~\cite{VWVE2014,GRVG2014,PVE2015,PCV2015,MRDPPR2016}, $E(\epsilon)$ is a non-monotonic function with a minimum at  the most probable value $\bar{\epsilon}$ and equal asymptotes for $\epsilon \to \pm \infty$ (the convergence to the asymptotic limit  is  slow, likely following a power law~\cite{PVE2015,PCV2015}). We observe that  the least probable value of $\epsilon$, corresponding to the maximum of  $E(\epsilon)$, is {\it negative}. This feature distinguishes the present ``information engine" from the stochastic heat engines studied in \cite{VWVE2014,GRVG2014,PVE2015,PCV2015,MRDPPR2016}.
}

Having determined the LDFs for information and heat, we now turn to the structure and origin of these fluctuations.
We begin by addressing the typical information required to produce a rare fluctuation of heat, or {the other way around}. In Fig. \ref{Figopt}, we plot  the most probable value of $\imath^v$ (resp. $\sigma^m$) for a given rare fluctuation of $\sigma^m$ (resp. $\imath^v$).  These quantities are computed from  the derivatives of the joint SCGF $\mu_{\sigma^m,\imath^v}(\lambda_1,\lambda_2)$ at $\lambda_2=0$ (resp. $\lambda_1=0$) (see SM) (note that we restrict our study to the range $\sigma^m<0.712$ where fluctuations of the boundary term are irrelevant). 
\begin{figure}[hbt]
 \begin{center}
\includegraphics[trim={0 1.25cm 0 1.75cm},clip,width=7cm]{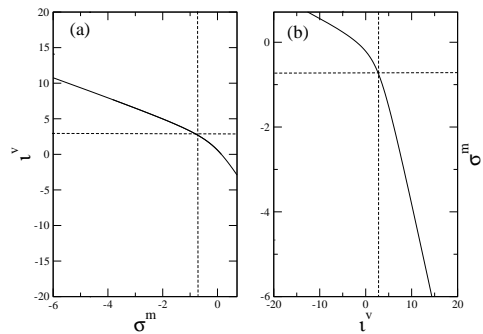}
\caption{(a) Most probable value of $\imath^v$  for a given value of $\sigma^m$. (b) Most probable value of  $\sigma^m$  for a given value of $\imath^v$. The dashed lines indicate the typical values $\langle \sigma^m\rangle\approx-0.72$ and $\langle \imath^v\rangle\approx 2.70$.}
\label{Figopt}
\end{center}
\end{figure}

As could be expected intuitively,  the fluctuations of $\sigma^m$ and $\imath^v$  are strongly correlated. But a less predictable and remarkable feature is the asymmetry between positive and negative fluctuations: observing a negative fluctuation $\sigma^m< \langle \sigma^m\rangle$ {(resp. a positive fluctuation $\imath^v>\langle \imath^v\rangle$)} for a long time requires a  smaller {(resp. larger)} variation of $\imath^v$ {(resp. $\sigma^m$)} than observing $\sigma^m> \langle \sigma^m\rangle$  {(resp. $\imath^v<\langle \imath^v\rangle$)}. 
 \begin{figure}
\begin{center}
\includegraphics[trim={0 1.25cm 0 1.75cm},clip,width=7cm]{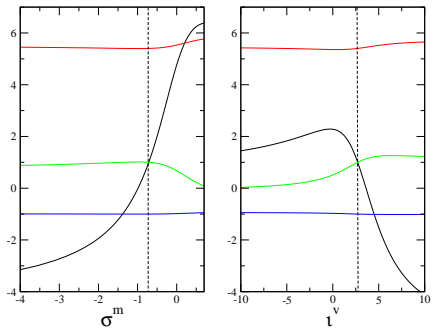}
\caption{ (Color on line) Effective interactions of the auxiliary processes that generate atypical values of $\sigma^m$ (left panel) or $\imath^v$ (right panel): $\gamma_{\rm eff}$ (black line), $\kappa_{\rm eff}$ (red line),  $k_1$ (blue line), $k_2$ (green line). Vertical dashed lines indicate the typical values of $\sigma^m$ and $\imath^v$.}
\label{FigForces}
\end{center}
\end{figure}

Deeper insight into the origin of these fluctuations is offered by studying the two auxiliary (or driven) processes~\cite{GL2010,JS2010,CT2013} that generate the constrained ensembles ${\cal P}[{\bf v}_0^t,{\bf y}_0^t\vert \Sigma^m=\sigma^m t]$ or ${\cal P}[{\bf v}_0^t,{\bf y}_0^t\vert I^v=\imath^v t]$ asymptotically. Rare fluctuations then become typical. These auxiliary dynamics are again linear (see SM for  details), with modified effective interactions
\begin{align}
\label{Eqeff} 
m\dot v_t&=-\gamma_{\rm eff} v_t -\kappa_{\rm eff}y_t+\xi_t\nonumber\\
\tau \dot y_t&=k_1 y_t+k_2 v_t+\eta_t\ .
\end{align}

The variations of the coefficients with $\sigma^m$ or $\imath^v$ are shown in Fig. \ref{FigForces}.
Again, we observe different behavior for negative and positive  fluctuations. The atypical events $\sigma^m< \langle \sigma^m\rangle$ or $\imath^v>\langle \imath^v\rangle$ are created essentially by a decrease of the friction coefficient, which even becomes negative ($\kappa_{\rm eff}$ and $k_2$ also vary, but slightly, and $k_1$ does not change).
As a result, the fluctuations of $v_t$ are enhanced, as confirmed by Fig. \ref{Pstdriven}, and the effective kinetic temperature increases. 
This additional uncertainty about $v_t$ allows for more information to be gathered, leading to the corresponding increase in $\imath^v$ observed in Fig. 3a. In other words, acquiring more information does not necessarily mean that it is effectively used to cool the system.
\begin{figure}[hbt]
\begin{center}
\includegraphics[trim={0 1.25cm 0 1.75cm},clip,width=7cm]{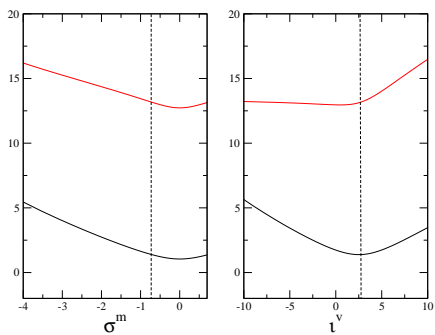}
\caption{ (Color on line) $\langle v^2_t\rangle$ (black line) and  $\langle y^2_t\rangle $ (red line)  conditioned on a given value of $\sigma^m$ (left) or $\imath^v$ (right).}
\label{Pstdriven}
\end{center}
\end{figure}

The case of atypical events $\sigma^m> \langle \sigma^m\rangle$ or $\imath^v<\langle \imath^v\rangle$  is not so simple, as both $\gamma_{\rm eff}$ and $k_2$ vary significantly. Moreover, $\gamma_{\rm eff}$ is not a monotonic function of $\imath^v$. Remarkably, $k_2$ becomes very small as $\imath^v$ becomes very negative. The demon then does not perform any measurement, and $y_t$ just plays the role of additional noise.
\begin{figure}[hbt]
\begin{center}
\includegraphics[trim={0 1.25cm 0 1.75cm},clip,width=7cm]{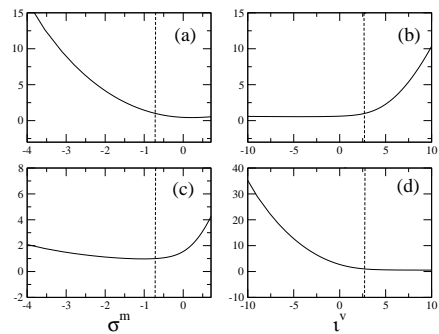}
\caption{Intensity of the atypical noises as a function of $\sigma^m$ or $\imath^v$.  (a) and (b): $S_{\xi_{\rm atyp}}(\omega=0)/(2\gamma T)$; (c) and (d): $S_{\eta_{\rm atyp}}(\omega=0)/\Delta$. Vertical dashed lines indicate the typical values of $\sigma^m$ and $\imath^v$.}
\label{FigAtyp}
\end{center}
\end{figure}

An alternative approach to understanding the origin of large deviations is to consider the atypical noise realizations that create rare fluctuations and determine whether fluctuations in $\xi_t$ or $\eta_t$ play the dominant role.  To this end, we select an atypical  trajectory $({\bf v}_0^t,{\bf y}_0^t)_{\rm atyp}$  produced by the auxiliary processes ({\ref{Eqeff}) for a given value of $\sigma^m$ or $\imath^v$ and insert it into the original equations of motion ({\ref{Eqdef}). This yields two biased noises $\xi_{\rm atyp}$ and $\eta_{\rm atyp}$. Thanks to the linearity of the equations, this calculation can be performed by working in the frequency domain, {which yields
\begin{align}
\xi_{\rm atyp}(\omega)&=(\gamma -im\omega)v_{\rm atyp}(\omega)+\kappa\: y_{\rm atyp}(\omega)\nonumber\\
&=\frac{\chi_{\rm eff}(\omega)}{k_1+i\omega \tau}\Big[ [(\gamma -im\omega)(k_1+i\omega \tau)-\kappa k_2]\xi(\omega)\nonumber\\
&+[(\gamma -im\omega)\kappa_{\rm eff}-(\gamma_{\rm eff}-im\omega)\kappa]\eta(\omega)\Big]\nonumber\\
\eta_{\rm atyp}(\omega)&=(1-i\omega \tau)y_{\rm atyp}(\omega)-v_{\rm atyp}(\omega)\nonumber\\
&=-\frac{\chi_{\rm eff}(\omega)}{k_1+i\omega \tau}\Big[[(k_1+i\omega \tau)+k_2(1-i\omega \tau)]\xi(\omega)\nonumber\\
&+[(\gamma_{\rm eff}-im\omega)(1-i\omega \tau)+\kappa_{\rm eff}]\eta(\omega)\Big]\ ,\end{align}
with 
\begin{align} 
\chi_{\rm eff}(\omega)=\frac{k_1+i\omega \tau}{(\gamma_{\rm eff}-im\omega)(k_1+i\omega \tau)-\kappa_{\rm eff}k_2}\ .
\end{align}
Therefore, the two noises are correlated and colored, with power spectral densities $S_{\xi_{\rm atyp}}(\omega)=\langle\xi_{\rm atyp}(\omega)\xi_{\rm atyp}(-\omega)\rangle$ and $S_{\eta_{\rm atyp}}(\omega)=\langle\eta_{\rm atyp}(\omega)\eta_{\rm atyp}(-\omega)\rangle$.}
For simplicity, we here only characterize $\xi_{\rm atyp}$ and $\eta_{\rm atyp}$ by their intensity, that is the zero-frequency part of their power spectrum. The variations of $S_{\xi_{\rm atyp}}(\omega=0)$ and $S_{\xi_{\rm atyp}}(\omega=0)$  (normalized by the intensities of the original white noises) as a function of $\sigma^m$ or $\imath^v$ are shown in Fig. \ref{FigAtyp}. The overall picture is  again very  instructive: atypical events $\sigma^m< \langle \sigma^m\rangle$ or $\imath^v>\langle \imath^v\rangle$ are mainly due to an atypical history of the thermal noise $\xi_t$ whereas  atypical events $\sigma^m>\langle \sigma^m\rangle$ or $\imath^v<\langle \imath^v\rangle$ are  mainly due to an atypical history of the measurement noise $\eta_t$. 

\section{Conclusion}
  In this Letter, we have studied a Brownian particle model of feedback cooling, as a realization of  an autonomous Maxwell's demon. We have first derived a series of fluctuation theorems for the information flow and the partial entropy production in coupled diffusion processes.
We then investigated the fluctuations of information flow and entropy flow, and their correlations, focusing on the long-time limit. By analyzing in detail the effective dynamics and atypical noise realizations that instigate rare fluctuations, we have unraveled the subtle trade-off between noise in the system, noise in the measurement device, and  control efficiency. We believe that using the same approach with other models of Maxwell's demon could significantly improve our understanding of the thermodynamics of information. 

\acknowledgments
We wish to thank the organizers of the workshop {\it New Frontiers in Non-equilibrium Physics} held at the Yukawa Institute of Theoretical Physics  (Kyoto, 2015) where this work was initiated. MLR also thanks T. Munakata for helpful discussions. JMH is supported by the Gordon and Betty Moore Foundation through Grant GBMF4343.

\pagebreak
\widetext
\begin{center}
\textbf{\large Supplemental Material}
\end{center}

\setcounter{section}{0}
\setcounter{equation}{0}
\setcounter{figure}{0}
\makeatletter
\renewcommand{\theequation}{S\arabic{equation}}
\renewcommand{\thefigure}{S\arabic{figure}}
\renewcommand{\bibnumfmt}[1]{[S#1]}
\renewcommand{\citenumfont}[1]{S#1}

\section{Fluctuation relations}

\subsection{IFT for $\Sigma$}

We first derive the IFT  for the (partial) entropy production  $\Sigma=\Sigma^v+I^v$.  As stated in the main text, this boils down to showing that
\begin{align} 
\label{EqDFTsigma}
\Sigma&=\ln \frac{{\cal P}[{\bf v}_0^t,{\bf y}_0^t]}{{\cal P}^*[\tilde {\bf v}_0^t,\tilde {\bf y}_0^t]}\ ,
\end{align}
where ${\cal P}^*[\tilde {\bf v}_0^t,\tilde {\bf y}_0^t]\equiv {\cal P}^*[\tilde {\bf v}_0^t\vert \tilde {\bf y}_0^t,\tilde v_0]{\cal P}^*[\tilde {\bf y}_0^t\vert \tilde{\bf  v}_0^t,\tilde y_0]p_t(v_t,y_t)$ is the joint probability density of the  time-reversed  path $(\tilde{\bf v}_0^t,\tilde{\bf y}_0^t)=(-{\bf v}_t^0,{\bf y}_t^0)$ generated by a modified process, hereafter denoted by the star symbol.  The probabilities ${\cal P}[{\bf v}_0^t\vert  {\bf y}_0^t, v_0]$, ${\cal P}[{\bf y}_0^t\vert  {\bf v}_0^t, y_0]$, etc. are expressed in terms of Onsager-Machlup (OM) action  functionals.

Following [8], we anticipate that the star dynamics only modifies the equation of motion for $y_t$, so that ${\cal P}^*[\tilde {\bf v}_0^t,\tilde {\bf y}_0^t]= {\cal P}[\tilde {\bf v}_0^t\vert \tilde {\bf y}_0^t,\tilde v_0]{\cal P}^*[\tilde {\bf y}_0^t\vert \tilde{\bf  v}_0^t,\tilde y_0]p_t(v_t,y_t)$. Therefore,
\begin{align} 
\label{}
 \ln\frac{{\cal P}[{\bf v}_0^t,{\bf y}_0^t]}{{\cal P}^*[\tilde {\bf v}_0^t,\tilde {\bf y}_0^t]}=\Sigma^m +\ln\frac{{\cal P}[ {\bf y}_0^t\vert {\bf  v}_0^t,y_0]p_0(v_0, y_0)}{{\cal P}^*[\tilde {\bf y}_0^t\vert \tilde{\bf  v}_0^t,\tilde y_0]p_t(v_t,  y_t)}\ ,
\end{align}
where we have used the  local detailed balance relation $\ln \big({\cal P}[ {\bf  v}_0^t\vert  {\bf y}_0^t,v_0]/{\cal P}[\tilde  {\bf v}_0^t\vert \tilde  {\bf y}_0^t,\tilde v_0]\big)=\Sigma^m$.
Introducing  the quantity $\dot s_t^v(v_t,y_t)$,  we see that eq.\ (\ref{EqDFTsigma}) is satisfied if 
\begin{align} 
\label{Eqpartial1}
\ln\frac{{\cal P}[ {\bf y}_0^t\vert {\bf  v}_0^t,y_0]}{{\cal P}^*[\tilde {\bf y}_0^t\vert \tilde{\bf  v}_0^t,\tilde y_0]}
&=\int_0^t dt'\: {\dot s}_{t'}^v(v_{t'},y_{t'})+\ln\frac{p_t(v_t, y_t)}{p_0(v_0, y_0)} \nonumber\\
&\equiv-\int_0^t dt'\: {\dot s}_{t'}^y(v_{t'},y_{t'})
\end{align}
where ${\dot s}_t^y(v_t,y_t)\equiv \frac{d}{dt}s_t(v_t,y_t)-{\dot s}_t^v(v_t,y_t)$ is the time-variation of the stochastic joint entropy $s_t(v_t, y_t)$ due to $y$'s dynamics. Hence
\begin{align} 
\label{Eqsy}
{\dot s}_t^y(v_t,y_t)=\frac{1}{p_t(v_t,y_t)}\partial_y J^y_t(v_t,y_t)-\dot y_t \partial_y \ln p_t(v_t,y_t)\ .
\end{align}
Thus, the derivation of the IFT rests on the construction of a Langevin process that generates trajectories ${\bf y}_0^t$ with a conditional weight ${\cal P}^*[\tilde {\bf y}_0^t\vert \tilde{\bf  v}_0^t,\tilde y_0]$ obeying eq.\ (\ref{Eqpartial1}). As we now show, this process is governed by the  equation of motion
\begin{align}
\label{Eqauxdef} 
\tau \dot y_t=v_t+y_t+\frac{\Delta}{\tau}\partial_y \ln p_t(-v_t,y_t)+\eta_t\ ,
\end{align}
where  $\eta_t$ is the same Gaussian white noise as in the original process.  Equation (\ref{Eqauxdef}) may be viewed as an overdamped Langevin equation  with a time-dependent force $F^*(v_t,y_t,t)=v_t+y_t+(\Delta/\tau)\partial_y \ln p_t(-v_t,y_t)$ and $\tau^{-1}$ playing the role of a mobility $\mu_y$. Hence 
\begin{align}
\label{EqOMy}
{\cal P}^*[\tilde {\bf y}_0^t\vert \tilde{\bf  v}_0^t,\tilde y_0]&\propto e^{-\frac{1}{2\Delta}\int_0^t dt'\:[\tau \dot y_{t'}-F^*(v_{t'},y_{t'},t')]^2-\frac{1}{2\tau}\int_0^tdt'\:\partial_y F(v_{t'},y_{t'},t')}\nonumber\\
&\propto e^{-\frac{t}{2\tau}}e^{-\frac{1}{2\Delta}\int_0^t dt'\:\big\{[\tau \dot y_{t'}-v_{t'}+y_{t'}+\frac{\Delta}{\tau}\partial_y \ln p_{t'}(v_{t'},y_{t'})]^2+(\frac{\Delta}{\tau})^2\partial_y^2 \ln p_{t'}(v_{t'},y_{t'})]\big\}}\ ,
\end{align}
where we have changed $t'$ into $-t'$ in  the second line of the equation (we recall that the velocity $v_t$ is odd under time reversal but that $y_t$ must be treated as an even variable~[16]).  Comparing with the conditional density associated with the original Langevin equation,
\begin{align}
{\cal P}[ {\bf y}_0^t\vert {\bf  v}_0^t,y_0]\propto e^{\frac{t}{2\tau}}e^{-\frac{1}{2\Delta}\int_0^t dt'\:[\tau \dot y_{t'}+y_{t'}-v_{t'}]^2}\ ,
\end{align}
we obtain
\begin{align}
\label{EqOMy1} 
\ln\frac{{\cal P}[ {\bf y}_0^t\vert {\bf  v}_0^t,y_0]}{{\cal P}^*[\tilde {\bf y}_0^t\vert \tilde{\bf  v}_0^t,\tilde y_0]}=\frac{t}{\tau}+\frac{1}{\tau}\int_0^t dt'\:\big\{(\tau \dot y_{t'}+y_{t'}-v_{t'})\partial_y \ln p_{t'}(v_{t'},y_{t'})+\frac{\Delta}{2\tau}\frac{\partial_y^2 p_{t'}(v_{t'},y_{t'})}{p_{t'}(v_{t'},y_{t'})}\big\}\ ,
\end{align}
where we have used the identity $(\partial_y \ln f)^2+\partial_y^2 \ln f=f^{-1}\partial_y^2 f$. On the other hand, by inserting  the expression of the probability current $J_t^y(v_t,y_t)=-\tau^{-1}(y_t-v_t)p(v_t,y_t)-\Delta/(2\tau^2)\partial_yp_t(v_t,y_t)$ into eq.\ (\ref{Eqsy}), we find
\begin{align} 
\dot s_t^y(v_t,y_t)=-\frac{1}{\tau}-\frac{1}{\tau}(y_t-v_t)\partial_y\ln p_t(v_t,y_t)-\frac{\Delta}{2\tau^2}\frac{\partial_y^2 \:p_t(v_t,y_t)}{p_t(v_t,y_t)}-\dot y_t\partial_y\ln p_t(v_t,y_t)\ .
\end{align}
Therefore eq.\ (\ref{Eqpartial1}) is  satisfied, as announced. 

As stated in the main text, the IFT  holds for any coupled Langevin processes involving independent noises, for instance the two  overdamped Langevin equations (17). Eq.\ (\ref{EqOMy1}) is then  replaced by
\begin{align} 
\label{EqOMy2} 
\ln\frac{{\cal P}[ {\bf y}_0^t\vert {\bf  x}_0^t,y_0]}{{\cal P}^*[\tilde {\bf y}_0^t\vert \tilde{\bf  x}_0^t,\tilde y_0]}&=\frac{1}{4T_y}\int_0^t dt'\: \Big[2\dot y_{t'}-\mu_y[F_2(x_{t'},y_{t'},t')-F_2^*(x_{t'},y_{t'},t')]\Big]\Big[F_2(x_{t'},y_{t'},t')+F_2^*(x_{t'},y_{t'},t')\Big]\nonumber\\
&-\frac{\mu_y}{2}\int_0^t dt'\: \partial_y [F_2(x_{t'},y_{t'},t')-F_2^*(x_{t'},y_{t'},t')]\ ,
\end{align}
where $F_2^*(x,y,t)$ is the force acting on $y_t$ in the modified dynamics. Then, by choosing 
\begin{align} 
F_2^*(x,y,t)=-F_2(x_t,y_t,t)+2T_y\partial_y\ln p_t(x_t,y_t)\ ,
\end{align}
the r.h.s. of eq.\ (\ref{EqOMy2})  identifies with $\int_0^t dt'\: \dot s_t^y(x_{t'},y_{t'})$, with $\dot s_t^y(v_t,y_t)$ defined by eq.\ (\ref{Eqsy}).  We leave the demonstration as an exercise for the reader. 

\subsection{IFT for $I^v$}

The IFT for the integrated  information flow  [eq.\ (21) in the main text] is obtained by modifying also the  dynamics of $v_t$ and changing $\gamma$ into  $-\gamma$ while keeping the variance of $\xi_t$ fixed (this modified process is denoted by the hat symbol hereafter). As shown in [28], this leads to
\begin{align}
\label{Eqratio}
\frac{{\cal P}[{\bf v}_0^t\vert {\bf y}_0^t,v_0]}{\hat{\cal P}[{\bf v}_0^t\vert {\bf y}_0^t,v_0]}=e^{\frac{\gamma}{m}t}e^{\Sigma^m} \ .
\end{align}
By replacing the trajectories by their time-reversed images, this relation can be  rewritten as
\begin{align}
\label{Eqratio1}
\Sigma^m=\frac{\gamma}{m}t+\ln \frac{\hat {\cal P}[\tilde {\bf v}_0^t\vert \tilde{\bf y}_0^t,\tilde v_0]}{{\cal P}[ \tilde {\bf v}_0^t\vert \tilde {\bf y}_0^t,\tilde v_0]}\ .
\end{align}
Therefore, using eq.\ (\ref{EqDFTsigma}), we have
\begin{align} 
\Sigma-\Sigma^m&=-\frac{\gamma}{m}t+\ln \frac{{\cal P}[{\bf v}_0^t,{\bf y}_0^t]}{{\cal P}^*[\tilde {\bf v}_0^t,\tilde {\bf y}_0^t]}-\ln \frac{\hat {\cal P}[\tilde {\bf v}_0^t\vert \tilde{\bf y}_0^t,\tilde v_0]}{{\cal P}[ \tilde {\bf v}_0^t\vert \tilde {\bf y}_0^t,\tilde v_0]}\nonumber\\
&=-\frac{\gamma}{m}t+\ln \frac{{\cal P}[{\bf v}_0^t,{\bf y}_0^t]}{{\cal P}[\tilde {\bf v}_0^t\vert\tilde {\bf y}_0^t,\tilde v_0]{\cal P}^*[\tilde {\bf y}_0^t\vert\tilde {\bf v}_0^t,\tilde y_0]p_t(v_t,y_t)}-\ln \frac{\hat {\cal P}[\tilde {\bf v}_0^t\vert \tilde{\bf y}_0^t,\tilde v_0]}{{\cal P}[ \tilde {\bf v}_0^t\vert \tilde {\bf y}_0^t,\tilde v_0]}\nonumber\\
&=-\frac{\gamma}{m}t+\ln \frac{{\cal P}[{\bf v}_0^t,{\bf y}_0^t]}{\hat {\cal P}^*[\tilde {\bf v}_0^t, \tilde{\bf y}_0^t]}\ ,
\end{align}
where $\hat {\cal P}^*[\tilde {\bf v}_0^t,\tilde{\bf y}_0^t]\equiv \hat {\cal P}[\tilde {\bf v}_0^t\vert \tilde{\bf y}_0^t,\tilde v_0]{\cal P}^*[\tilde {\bf y}_0^t\vert\tilde {\bf v}_0^t,\tilde y_0]p_t(v_t,y_t)$ is the joint probability density of the backward path generated by the modified  ``hat-star" processs (i.e., the ``hat" dynamics for $v_t$ and the ``star" dynamics for $y_t$). Since $\Sigma-\Sigma^m=I^v-\ln p_t(v_t)/p_0(v_0)$ (cf. eqs. (8) and (9) in the main text), we finally obtain
\begin{align} 
\label{EqPhatstar3}
\frac{p_t(v_t)}{p_0(v_0)}e^{-I^v}&=e^{\frac{\gamma}{m}t}\frac{\hat {\cal P}^*[\tilde {\bf v}_0^t, \tilde{\bf y}_0^t]}{{\cal P}[{\bf v}_0^t,{\bf y}_0^t]}\ ,
\end{align}
which leads to the IFT for the  information flow by integration over the ensemble of paths generated by the original process. Note that this IFT is less general than the one for $\Sigma$ since the force acting on the state variable of the first subsystem must be linear. This is true for the model under study since this force is just the viscous force $-\gamma v_t$. For the coupled overdamped Langevin processes described by eqs. (17), one must have $F_1(x,y,t)=-kx_t +F(y_t,t)$. Then $(\gamma/m)t$ is replaced by $(\mu_x k)t$ in eqs. (\ref{Eqratio})-(\ref{EqPhatstar3}) (see [28] for a more general discussion).

\section{Scaled cumulant generating functions}

In this section we derive  eqs. (27)  in the main text, and we discuss the domain of validity of these expressions. We  also give the expressions of $\mu_{\sigma}(\lambda)$ and of the joint SCGF $\mu_{\sigma^m, \imath^v}(\lambda_1,\lambda_2)$. These results are obtained by imposing periodic boundary conditions on the trajectories and expanding $v_t$ and $y_t$ in discrete Fourier series,
\begin{align} 
v(t)&=\sum_{n=-\infty}^{\infty} v_ne^{-i\omega_n t}\nonumber\\
y(t)&=\sum_{n=-\infty}^{\infty} y_ne^{-i\omega_n t}\ ,
\end{align}
with  inverse transforms
\begin{align} 
v_n&=\frac{1}{t} \int_0^t ds \:v(s) e^{i\omega_n s}\nonumber\\
y_n&=\frac{1}{t} \int_0^t ds \:y(s) e^{i\omega_n s}\ ,
\end{align}
where $\omega_n=2\pi n/t$ and $v_n\equiv v(\omega_n)$, $y_n\equiv y(\omega_n)$.   
 The coupled Langevin equations in the frequency domain are then
\begin{align} 
(\gamma -im\omega_n)v_n&=-\kappa y_n+\xi_n\nonumber\\
(1-i\omega_n\tau)y_n&=v_n+\eta_n\ ,
\end{align}
with $\langle \xi_n \xi_{n'}\rangle=(2\gamma T/t)\delta_{n,-n'}$, $\langle \eta_n\eta_{n'}\rangle=(\Delta/t)\delta_{n,-n'}$, and  $\langle \xi_n\eta_{n'}\rangle=0$.
This leads to
\begin{align}
\label{EqL} 
v_n&=\chi_n[\xi_n-\frac{\kappa }{1-i\omega_n \tau}\eta_n]\nonumber\\
y_n&=\frac{\chi_n}{1-i\omega_n \tau}[\xi_n+(\gamma-im\omega_n)\eta_n]
\end{align}
where 
\begin{align} 
\chi_n\equiv \chi(\omega_n)=\frac{1-i\omega_n \tau}{\kappa+\gamma-i(m+\gamma \tau)\omega_n -m\omega_n^2\tau}\ .
\end{align}

\subsection{SCGFs for single observables}\label{sec:SCGF}

For brevity,  we will only detail the calculation of $\mu_{\imath^v}(\lambda)$. In the stationary state, $I^v$ is given by eq.\ (23) in the main text. We recall that $\alpha_{11},\alpha_{22}$ and $\alpha_{12}=\alpha_{21}$ are the entries of ${\bf C}^{-1}$, the inverse of the covariance matrix whose expression is given in Appendix D of [16].
In terms of the Fourier coefficients $v_n$ and $y_n$,  the rate $\imath^v\equiv I^v/t$  reads 
\begin{align} 
\label{Eqiv1}
\imath^v&=-A_1+\sum_{n=-\infty}^{\infty}[\alpha_{11}A_1v_n v_{-n}+A_2y_ny_{-n}+(A_3-i\omega_n\alpha_{12})v_ny_{-n}]+b.t.
\end{align}
where $b.t.$ is a temporal boundary term that is neglected hereafter (see the discussion below). Inserting eqs. (\ref{EqL})  then yields 
\begin{align} 
\label{Eqiv2}
\imath^v &\sim -A_1+\sum_{n=-\infty}^{\infty} \frac{\vert \chi_n\vert^2}{1+\omega_n^2\tau^2}\Big\{[\alpha_{11}A_1(1+\omega_n^2\tau^2)+A_2+(A_3-i\omega_n\alpha_{12})(1-i\omega_n\tau)]\vert \xi_n\vert^2\nonumber\\
&+[\kappa^2\alpha_{11}A_1+(\gamma^2+m^2\omega_n^2)A_2-\kappa(A_3-i\omega_n\alpha_{12})(\gamma+im\omega_n)]\vert \eta_n\vert^2\nonumber\\
&+[-\kappa \alpha_{11}A_1(1-i\omega_n\tau)+(\gamma+im\omega_n)A_2+(A_3-i\omega_n\alpha_{12})(\gamma+im\omega_n)(1-i\omega_n\tau)]\xi_n\eta_{-n} \nonumber\\
&+[-\kappa\alpha_{11}A_1(1+i\omega_n\tau)+(\gamma-im\omega_n)A_2-\kappa(A_3-i\omega_n\alpha_{12})]\xi_{-n}\eta_n\Big\}\ , 
\end{align}
which is rewritten in a compact form as
\begin{align} 
\label{Eqiv3}
\imath^v &\sim -A_1+\sum_{n=1}^{\infty} \zeta_n^T{\bf L}_{n,\imath^v}\zeta^*_n\ ,
\end{align}
where $\zeta_n=(\xi_n,\eta_n)^T$, the symbol $*$ denotes the complex conjugate (with $\xi^*_n=\xi_{-n},\eta^*_n=\eta_{-n}$), and ${\bf L}_{n,\imath^v}$ is a $2\times 2$ Hermitian matrix with entries 
\begin{align} 
\label{EqLi}
L_{11,\imath^v}(\omega_n)&=\frac{2\vert \chi_n\vert^2}{1+\omega_n^2\tau^2}[\alpha_{11}A_1(1+\omega_n^2\tau^2)+A_2+A_3-\omega_n^2\tau\alpha_{12}]\nonumber\\
L_{12,\imath^v}(\omega_n)&=L^*_{21}(\omega_n)=\frac{\vert \chi_n\vert^2}{1+\omega_n^2\tau^2}\Big[-2\kappa\alpha_{11}A_1(1-i\omega_n\tau)+2(\gamma+im\omega_n)A_2+A_3[(\gamma+im\omega_n)(1-i\omega_n\tau)-\kappa]\nonumber\\
&-i\omega_n\alpha_{12}[(\gamma+im\omega_n)(1-i\omega_n\tau)+\kappa]\Big]\nonumber\\
L_{22,\imath^v}(\omega_n)&=\frac{2\vert \chi_n\vert^2}{1+\omega_n^2\tau^2}[\kappa^2\alpha_{11}A_1+(\gamma^2+m^2\omega_n^2)A_2-\kappa(A_3\gamma+m\omega_n^2\alpha_{12})]\ .
\end{align}
This leads to
\begin{align} 
\label{EqZn}
 \langle e^{-\lambda I^v}\rangle&\sim e^{\lambda A_1t} 
\prod_{n=1}^{\infty}\int d\zeta_n\: P(\zeta_n)e^{-\lambda t\:\zeta_n^T{\bf L}_{n,\imath^v}\zeta^*_n}\ ,
\end{align}
where
\begin{align} 
P(\zeta_n)=\frac{1}{\pi^2\mbox{det}{\bf D}}e^{-\zeta_n{\bf D}^{-1}\zeta_n^*}
\end{align}
and 
\[
{\bf D}=\frac{1}{t}\left(
\begin{array}{cccc}
 2\gamma T &0 \\
0& \Delta&
\end{array}
\right) \ .
\]
The Gaussian integration over $\zeta_n$  then gives
\begin{align} 
\int d\zeta_n\: P(\zeta_n)e^{-\lambda t\:\zeta_n^T{\bf L}_{n,\imath^v}\zeta^*_n}=\mbox{det}[{\bf I}+\lambda t{\bf D}{\bf L}_{n,\imath^v}]^{-1}\ .
\end{align}
In the long-time limit, the summation over $n$ can be replaced by an integral over $\omega$, and we finally obtain
\begin{align} 
\label{Eqmuiv}
\mu_{\imath^v}(\lambda)&=\frac{\gamma}{m}(1-T\frac{\alpha_{11}}{m})\lambda-\int_{0}^{\infty}\frac{d\omega}{2\pi} \ln [1-F_{\imath^v,\lambda}(\omega)]\ ,
\end{align}
with 
\begin{align} 
\label{EqFi}
F_{\imath^v,\lambda}(\omega)=-2\gamma T \Delta \lambda^2[L_{11,\imath^v}(\omega)L_{22,\imath^v}(\omega)-L_{12,\imath^v}(\omega)L_{21,\imath^v}(\omega)]\ ,
\end{align}
which leads to eq.\ (27b) in the main text.

The calculation of $\mu_{\sigma^m}(\lambda)$ is quite similar but somewhat simpler since $\Sigma^m=-(\kappa/T)\int_0^t dt'\: v_{t'}y_{t'}\sim -(\kappa/T)\sum_n v_ny_n$.  The entries of the corresponding matrix ${\bf L}_{n,\sigma^m}$  are then given by
\begin{align} 
\label{EqLsigmam}
L_{11,\sigma^m}(\omega_n)&=-\frac{2\kappa}{T}\frac{\vert \chi_n\vert^2}{1+\omega_n^2\tau^2}\nonumber\\
L_{12,\sigma^m}(\omega_n)&=L^*_{21}(\omega_n)=-\frac{\kappa}{T}\frac{\vert \chi_n\vert^2}{1+\omega_n^2\tau^2}[(\gamma+im\omega_n)(1-i\omega_n\tau)-\kappa]\nonumber\\
L_{22,\sigma^m}(\omega_n)&=\frac{2\kappa^2\gamma}{T}\frac{\vert \chi_n\vert^2}{1+\omega_n^2\tau^2}\ ,
\end{align}
so that
\begin{align} 
\label{Eqmusigmam}
\mu_{\sigma^m}(\lambda)&=-\int_{0}^{\infty}\frac{d\omega}{2\pi} \ln [1-F_{\sigma^m,\lambda}(\omega)]\ ,
\end{align}
with 
\begin{align} 
\label{EqFsigmam}
F_{\sigma^m,\lambda}(\omega)=-\lambda[2\gamma T L_{11,\sigma^m}(\omega)+\Delta L_{22,\sigma^m}(\omega)]-2\gamma T \Delta \lambda^2[L_{11,\sigma^m}(\omega)L_{22,\sigma^m}(\omega)-L_{12,\sigma^m}(\omega)L_{21,\sigma^m}(\omega)]\ ,
\end{align}
which leads to eq.\ (27a) in the main text. The main difference with $F_{\imath^v,\lambda}(\omega)$ is that the  term  linear in  $\lambda$ does not vanish so that $F_{\sigma^m,\lambda}(\omega)$ is not an even function of $\lambda$.

Finally, the expression of  $\mu_{\sigma}(\lambda)$ is obtained by exploiting the fact that $\Sigma\sim \Sigma^m+I^v$ as $t \to \infty$. Hence 
\begin{align} 
\label{Eqmusigma}
\mu_{\sigma}(\lambda)&=\frac{\gamma}{m}(1-T\frac{\alpha_{11}}{m})\lambda-\int_{0}^{\infty}\frac{d\omega}{2\pi} \ln [1-F_{\sigma,\lambda}(\omega)]\ ,
\end{align}
with 
\begin{align} 
\label{EqFsigma}
F_{\sigma,\lambda}(\omega)=-\lambda[2\gamma T (L_{11,\sigma}(\omega)+\Delta L_{22,\sigma}(\omega)]-2\gamma T \Delta \lambda^2[L_{11,\sigma}(\omega)L_{22,\sigma}(\omega)-L_{12,\sigma}(\omega)L_{21,\sigma}(\omega)]
\end{align}
and $L_{ij,\sigma}=L_{ij,\sigma^m}+L_{ij,\imath^v}$. 

Note that the  SCGFs are real quantities in an open domain  $(\lambda_-,\lambda_+)$ (different for each function) in which the argument of the logarithm stays positive for all values of $\omega$. For instance, with the choice $\tau/\tau^v=0.01,\ \mbox{SNR}=40,\ g=g_{\rm opt}\approx 5.403$ for the model parameters, we find that $\mu_{\imath^v}(\lambda)$, $\mu_{\sigma^m}(\lambda)$, and $\mu_{\sigma}(\lambda)$ are defined in the intervals $-1.004<\lambda<1.004$, $-8.130<\lambda<1.727$, and $-1.931<\lambda<1.027$, respectively. The slopes of the SCGFs diverge at the boundaries, which implies that the corresponding Legendre transforms are asymptotically linear~[29].  However, this is no longer true if the pre-exponential factors (see eq.\ (25) in the main text and eq.\ (\ref{Eqgiv}) below) have pole singularities inside the domain of definition. These singularities result from rare but large fluctuations of  the boundary terms neglected in the preceding calculation, and  the leading contribution to the LDF then comes from the singularity (whose position fixes the slope of the LDF).  For instance, $g_{\sigma^m}(\lambda)$ diverges for $\lambda=\lambda_0\approx -6.39$, so that $E(\sigma^m) \approx 0.038+6.39\sigma^m$ for $\sigma^m>-\mu'_{\sigma^m}(\lambda_0)\approx 0.712$ (see Fig. 2 of the main text).

\subsection{Joint SCGFs}

The same method based on discrete Fourier transforms can be used to compute  joint SCGFs such as   $\mu_{\sigma^m,\imath^v}(\lambda_1,\lambda_2)\equiv \lim_{t \to \infty}(1/t)\ln \langle e^{-\lambda_1\Sigma^m-\lambda_2 I^v} \rangle$. Moreover, since $\Sigma\sim \Sigma^m+I^v$ in the long-time limit, the three functions $\mu_{\sigma^m,\imath^v}(\lambda_1,\lambda_2), \mu_{\sigma^m,\sigma}(\lambda_1,\lambda_2), \mu_{\sigma,\imath^v}(\lambda_1,\lambda_2)$ are not independent.
Specifically,  $\mu_{\sigma,\imath^v}(\lambda_1,\lambda_2)=\mu_{\sigma^m,\imath^v}(\lambda_1,\lambda_1+\lambda_2)$ and $\mu_{\sigma^m,\sigma}(\lambda_1,\lambda_2)=\mu_{\sigma^m,\imath^v}(\lambda_1+\lambda_2,\lambda_2)$, with
\begin{align} 
\mu_{\sigma^m,\imath^v}(\lambda_1,\lambda_2)=\frac{\gamma}{m}(1-T\frac{\alpha_{11}}{m})\lambda_2-\frac{1}{2}\int_{-\infty}^{\infty} \frac{d\omega}{2\pi}\: \ln [1-F_{\sigma^m,\imath^v,\lambda_1,\lambda_2}(\omega)]
\end{align}
and
\begin{align} 
F_{\sigma^m,\imath^v,\lambda_1,\lambda_2}(\omega)&=-2\gamma T[\lambda_1L_{11,\sigma^m}(\omega)+\lambda_2 L_{11,\imath^v}(\omega)]-\Delta[\lambda_1L_{22,\sigma^m}(\omega)+\lambda_2 L_{22,\imath^v}(\omega)]\nonumber\\
&-2\gamma T\Delta\Big\{[\lambda_1L_{11,\sigma^m}(\omega)+\lambda_2L_{11,\imath^v}(\omega)][\lambda_1  L_{22,\sigma^m}(\omega)+\lambda_2 L_{22,\imath^v}(\omega)]\nonumber\\
&-[\lambda_1 L_{12,\sigma^m}(\omega)+\lambda_2 L_{12,\imath^v}(\omega)][\lambda_1L_{12,\sigma^m}^*(\omega)+\lambda_2 L_{12,\imath^v}^*(\omega)]\Big\}\ .
\end{align}
One can readily check that $\mu_{\sigma^m}(\lambda)=\mu_{\sigma^m,\imath^v}(\lambda,0)$, $\mu_{\sigma}(\lambda)=\mu_{\sigma^m,\imath^v}(0,\lambda)$, and $\mu_{\sigma}(\lambda)=\mu_{\sigma^m,\imath^v}(\lambda,\lambda)$.

\section{Tilted and auxiliary processes}

In this section, we construct the so-called auxiliary (or driven) processes~[24-26] that describe how large fluctuations of  $\Sigma^m$,  $I^v$, or $\Sigma$  are created in the long-time limit. For each of these observables, we first determine the  so-called tilted generator and compute the dominant eigenvalue and the associated left and right eigenfunctions (the dominant eigenvalue identifies with the SCGF already obtained in section \ref{sec:SCGF}, but the knowledge of  the eigenfunctions allows us to compute the pre-exponential factors). We then determine the biased forces or biased noises that make a large deviation of the observable typical.

For brevity, we mostly focus on large deviations of $I^v$. We also closely follow the analysis and notations of [26].

\subsection{Spectral elements}

\subsubsection{Tilted generators}

From the definition of $I^v$ [eqs. (7) and (8) in the main text], we  note that this quantity belongs to the general class of trajectory observables of the form~[26]
\begin{align} 
I^v=\int_0^t f_{\imath^v}({\bf X}_t)dt'+ \int_0^t {\bf g}_{\imath^v}({\bf X}_{t'})\circ d{\bf X}_{t'}\ ,
\end{align}
where ${\bf X}_t=(v_t,y_t)$, $f_{\imath^v}(v,y)=\partial_v J^v(v,y)/p(v,y)$ (since $\partial_v J(v)=0$ in the stationary state), and ${\bf g}_{\imath^v}(v,y)$ is a two-dimensional vector function with components $(-\partial_v \ln p(y\vert v), 0)$. We then introduce the non-conservative process associated with the exponentially tilted trajectory ensemble
\begin{align} 
\label{Eqtildedpath}
{\cal P}_{\imath^v,\lambda}[{\bf v}_0^t,{\bf y}_0^t]\equiv \frac{e^{-\lambda I^v}{\cal P}[{\bf v}_0^t,{\bf y}_0^t] }{\langle e^{-\lambda I^v}\rangle}
\end{align}
that becomes equivalent in the limit $t \to \infty$ to the ensemble of trajectories conditioned on a particular value of $\imath^v$ (the equivalence holds because the LDF $E(\imath^v)$ is convex, and the value of $\lambda$  achieving the equivalence is then given by $\lambda = -E'(\imath^v)$).  
The corresponding tilted generator is given by~[26] 
\begin{align} 
\label{Eqtiltedgen}
{\cal L}_{\imath^v\lambda}={\bf F}(\nabla -\lambda {\bf g}_{\imath^v})+(\nabla -\lambda{\bf  g}_{\imath^v})\frac{{\bf D}'}{2}(\nabla -\lambda {\bf g}_{\imath^v})-\lambda f_{\imath^v}\ ,
\end{align}
where ${\bf F}$ is the two-dimensional drift  of the original process with components $(-(\gamma v+\kappa y)/m,(v-y)/\tau)$ and the noise covariance matrix ${\bf D}'$ is here defined as
\[
{\bf D}'=\left(
\begin{array}{cccc}
 \frac{2\gamma T}{m^2} &0 \\
0& \frac{\Delta}{\tau^2}&
\end{array}
\right) \ .
\]
The generating function ${\cal Z}_{\imath^v,\lambda}(v,y,t)=\langle e^{-\lambda I^v}\delta(v-v_t)\delta(y-y_t)\rangle$ (where the average is taken over all trajectories of duration $t$ ending at $(v,y)$ with initial state $(v_0,y_0)$ drawn from the stationary pdf $p(v_0,y_0)$)  evolves according to 
\begin{align} 
\label{EqFPZlambda}
\partial_t Z_{\imath^v,\lambda}(v,y,t)={\cal L}_{\imath^v,\lambda}^{\dag} Z_{\imath^v,\lambda}(v,y,t)\ ,
\end{align}
where ${\cal L}_{\imath^v,\lambda}^{\dag}$ is the dual of ${\cal L}_{\imath^v,\lambda}$. Namely,
\begin{align} 
\label{EqLitilteddag}
{\cal L}_{\imath^v,\lambda}^{\dag} &=\frac{\gamma T}{m^2} \frac{\partial^2}{\partial v^2}+\frac{\Delta}{2\tau^2}\frac{\partial^2}{\partial y^2}+\frac{1}{m} \left[\gamma v+\kappa y-2\frac{\lambda \gamma T}{m}\partial_v \ln p(y\vert v)\right]\partial_v+\frac{y-v}{\tau}\partial_y\nonumber\\
&+\frac{\lambda}{m} \Big[\gamma+(\gamma v+\kappa y)\partial_v \ln p(v)\Big] +\frac{\lambda \gamma T}{m^2}\Big[(\lambda+1)(\partial_v\ln p(y\vert v))^2+2\partial_v\ln p(y\vert v)\partial_v \ln p(v)\nonumber\\
&+(\partial_v \ln p(v))^2+\frac{\partial^2}{\partial v^2}\ln p(v)\Big]+\frac{\gamma}{m}+\frac{1}{\tau}\ ,
\end{align}
with $\partial_v \ln p(y\vert v)=-(\alpha_{11}-c_{11}^{-1})v-\alpha_{12}y$ and $\partial_v \ln p(v)=-c_{11}^{-1} v$ (we recall that $c_{ij}$ and $\alpha_{ij}$ are the entries of the covariance matrix and its inverse, respectively). 

In order to determine the asymptotic behavior of $Z_{\imath^v,\lambda}(v,y,t)$, we do not need to solve the spectral problem for ${\cal L}_{\imath^v,\lambda}$ and  ${\cal L}^{\dag}_{\imath^v,\lambda}$ in full generality but only to compute the dominant eigenvalue $\mu_{\imath^v}(\lambda)$ and the associated right and left eigenfunctions, solutions of the equations
\begin{align} 
\label{EqSP}
{\cal L}_{\imath^v,\lambda}\,  r_{\imath^v,\lambda}(v,y)=\mu_{\imath^v}(\lambda)\, r_{\imath^v,\lambda}(v,y)\nonumber\\
{\cal L}_{\imath^v,\lambda}^{\dag}\, l_{\imath^v,\lambda}(v,y)=\mu_{\imath^v}(\lambda)\, l_{\imath^v,\lambda}(v,y)\ ,
\end{align}
with normalization conditions $\int dv\: dy \:l_{\imath^v,\lambda}(v,y)=1$ and $\int dv\: dy\: r_{\imath^v,\lambda}(v,y)l_{\imath^v,\lambda}(v,y)=1$.  From eq.\ (\ref{EqLitilteddag}) and the corresponding expression of ${\cal L}_{\imath^v\lambda}$, we find that the bivariate Gaussian functions $r_{\imath^v,\lambda}(v,y)=e^{-(1/2)[A_{\imath^v}(\lambda)v^2+B_{\imath^v}(\lambda)y^2+2C_{\imath^v}(\lambda)yv]}$ and $l_{\imath^v,\lambda}(v,y)=e^{-(1/2)[A_{\imath^v}^{\dag}(\lambda)v^2+B_{\imath^v}^{\dag}(\lambda)y^2+2C_{\imath^v}^{\dag}(\lambda)yv]}$ are solutions (not yet normalized) of these equations,  with the coefficients $A_{\imath^v}(\lambda), A_{\imath^v}^{\dag}(\lambda)$, etc. obeying complicated algebraic equations. $\mu_{\imath^v}(\lambda)$ is eventually obtained as the solution of an algebraic equation of degree $6$, and the proper root is selected by imposing numerical agreement with eq.\ (\ref{Eqmuiv}).
Then, assuming the existence of a gap between $\mu_{\imath^v}$ and the first sub-dominant eigenvalue,   $Z_{\imath^v,\lambda}(v,y,t)$ takes the asymptotic form~[26]
\begin{align} 
Z_{\imath^v,\lambda}(v,y,t)\sim e^{\mu_{\imath^v}(\lambda)t} \int dv_0 dy_0 \:p(v_0,y_0)r_{\imath^v,\lambda}(v_0,y_0)\: l_{\imath^v,\lambda}(v,y) \ ,
\end{align}
so that $Z_{\imath^v}(\lambda,t)=\int dv\: dy\:Z_{\imath^v,\lambda}(v,y,t) \sim g_{\imath^v}(\lambda)e^{\mu_{\imath^v}(\lambda)t}$ with
\begin{align} 
\label{Eqgiv}
g_{\imath^v}(\lambda)=\int dv_0 dy_0 \:p(v_0,y_0)r_{\imath^v,\lambda}(v_0,y_0)\ .
\end{align}
Depending on the model parameters (e.g. the feedback gain), the above integral may diverge for certain values of $\lambda$, signaling that the fluctuations of the temporal boundary terms must not be neglected, as discussed at the end of section B.1. 

Similar calculations are performed for $\Sigma^m$ and $\Sigma$. For completeness, we report the corresponding expressions of the tilted dual generators ${\cal L}_{\sigma^m,\lambda}^{\dag}$ and ${\cal L}_{\sigma,\lambda}^{\dag}$:
\begin{align} 
\label{EqLdagsigmam}
{\cal L}_{\sigma^m,\lambda}^{\dag} &=\frac{\gamma T}{m^2} \frac{\partial^2}{\partial v^2}+\frac{\Delta}{2\tau^2}\frac{\partial^2}{\partial y^2}+\frac{1}{m} \left[(1-2\lambda)\gamma v+\kappa y\right]\partial_v+\frac{y-v}{\tau}\partial_y\nonumber\\
&+\lambda \left[(\lambda-1)\gamma \frac{v^2}{T}-\frac{\gamma}{m} \right]+\frac{\gamma}{m}+\frac{1}{\tau}\ ,
\end{align}
and
\begin{align} 
\label{EqLdagsigma}
{\cal L}_{\sigma,\lambda}^{\dag} &=\frac{\gamma T}{m^2} \frac{\partial^2}{\partial v^2}+\frac{\Delta}{2\tau^2}\frac{\partial^2}{\partial y^2}+\frac{1}{m} \left[(1-2\lambda)\gamma v+\kappa y-2\frac{\lambda \gamma T}{m}\partial_v \ln p(v,y)\right]\partial_v+\frac{y-v}{\tau}\partial_y\nonumber\\
&+\lambda \left[\gamma v[(\lambda-1)\frac{v}{T}+\frac{2\lambda}{m}\partial_v \ln p(v,y)]+(\lambda+1)\frac{\gamma T}{m^2}(\partial_v\ln p(v,y))^2\right]+\frac{\gamma}{m}+\frac{1}{\tau}\ .
\end{align}
Note  that ${\cal L}_{\sigma^m,\lambda=1}^{\dag} -{\cal L}_{\sigma^m,\lambda=0}^{\dag}=-2(\gamma/m)v\partial_v -(\gamma/m)
$ so that $\partial_tZ_{\sigma^m}(1,t)=(\gamma/m)Z_{\sigma^m}(1,t)$ by integration over $v$ and $y$, in agreement with the IFT $ Z_{\sigma^m}(1,t)=\langle e^{-\Sigma^m}\rangle =e^{(\gamma/m)t}$.
Likewise, ${\cal L}_{\sigma,\lambda=1}^{\dag} -{\cal L}_{\sigma,\lambda=0}^{\dag}=2(\gamma/m)v+(T/m)\partial_v \ln p(v,y)][\partial_v \ln p(v,y)-\partial_v]$ so that $p(v,y)$ is an eigenfunction of ${\cal L}_{\sigma,\lambda=1}^{\dag} $ associated with the eigenvalue $0$, in agreement with the IFT $\langle e^{-\Sigma}\rangle=1$. However, we  stress  that the two IFTs for $\Sigma^m$ and $\Sigma$ are valid at any time, and not only asymptotically.

\subsubsection{Small-$t$ expansion of $ Z_{\imath^v}(\lambda,t)$}

In order to investigate whether the symmetry of the SCGF $\mu_{\imath^v}(\lambda)$ [eq.\ (\ref{Eqmuiv}) above] reflects a more general  symmetry of the pdf for the information flow, we introduce the modified  function $\tilde Z_{\imath^v,\lambda}(v,y,t)=Z_{\imath^v,\lambda}(v,y,t)e^{\lambda \langle \imath^v\rangle t}$ which evolves according to 
\begin{align} 
\label{EqFPZlambdatilde}
\partial_t \tilde Z_{\imath^v,\lambda}(v,y,t)=\tilde {\cal L}_{\imath^v,\lambda}^\dag \tilde Z_{\imath^v,\lambda}(v,y,t)\ ,
\end{align}
with $\tilde {\cal L}_{\imath^v,\lambda}^\dag= {\cal L}_{\imath^v,\lambda}^\dag+\lambda \langle \imath^v\rangle$. The solution can be formally expanded in powers of $t$ as
\begin{align} 
\tilde Z_{\imath^v,\lambda}(v,y,t)=p(v,y)+ \sum_{n=1}^{\infty}\frac{1}{n!}\tilde Z_{\imath^v,\lambda}^{(n)}(v,y)t^n
\end{align}
with $\tilde Z_{\imath^v,\lambda}^{(n)}(v,y)=(\tilde {\cal L}_{\imath^v,\lambda}^\dag)^n p(v,y)$. For brevity, we here only report the expression of the term proportional to $t$,
\begin{align} 
\tilde Z_{\imath^v,\lambda}^{(1)}(v,y)=p(v,y)[f_0(v,y)+f_1(v,y)\lambda +f_2(v,y)\lambda^2 ]\ ,
\end{align}
with
\begin{align} 
f_0(v,y)&=\frac{1}{2\tau^2m^2(c_{11}c_{22}-c_{12}^2)^2}\Big[[2\gamma T\tau^2 c_{22}^2+\Delta m^2c_{12}^2-2m\tau (c_{11}c_{22}-c_{12}^2)(\gamma \tau c_{22}+mc_{12})]v^2\nonumber\\
&+[2\gamma T \tau^2 c_{12}^2+\Delta m^2c_{11}^2+2m\tau (c_{11}c_{22}-c_{12}^2)(\kappa\tau c_{12}-mc_{11})]y^2\nonumber\\
&-2[2\gamma T\tau^2 c_{12}c_{22}+\Delta m^2c_{11}c_{12}-m\tau  (c_{11}c_{22}-c_{12}^2)(c_{12}(\gamma \tau+m)+mc_{11}-\kappa\tau c_{22})]vy\nonumber\\
&-(c_{11}c_{22}-c_{12}^2)(2\gamma T\tau^2 c_{22}+\Delta m^2c_{11}-2m \tau (c_{11}c_{22}-c_{12}^2)(\gamma \tau+m))\Big]\nonumber\\
f_1(v,y)&=\frac{\gamma}{m^2 c_{11}(c_{11}c_{22}-c_{12}^2)^2}\Big[[T c_{22}(c_{11}c_{22}-2c_{12}^2)-m(c_{11}c_{22}-c_{12}^2)^2]v^2-T c_{12}^2c_{11}y^2\nonumber\\
&+[2T c_{12}^3-\frac{\kappa m}{\gamma}(c_{11}c_{22}-c_{12}^2)^2]vy+T c_{12}^2(c_{11}c_{22}-c_{12}^2)\Big]\nonumber\\
f_2(v,y)&=\frac{\gamma T c_{12}^2}{m^2c_{11}^2(c_{11}c_{22}-c_{12}^2)^2}(c_{12}v-c_{11}y)^2\ .
\end{align}
Remarkably, the term linear in $\lambda$ cancels by integration over $v$ and $y$, and we obtain 
\begin{align} 
\tilde Z_{\imath^v}(\lambda,t)\equiv Z_{\imath^v}(\lambda,t)e^{\lambda \langle \imath^v\rangle t} =1+\frac{\gamma T}{m^2}\frac{ c_{12}^2}{c_{11}(c_{11}c_{22}-c_{12}^2)}\lambda^2t +{\cal O}(t^2) \ .
\end{align}
The calculation of higher-order terms quickly becomes  cumbersome  and we have been able to perform the expansion of $\tilde Z_{\imath^v,\lambda}(v,y,t)$ up to order $t^3$ only. Again, we find, after integrating over $v$ and $y$, that odd powers of $\lambda$ do not contribute to $\tilde Z_{\imath^v}(\lambda,t)$. This leads us  to conjecture that $\tilde Z_{\imath^v}(\lambda,t)$ is an even function of $\lambda$.

We have performed the same calculation for a non-linear model  with a feedback force $f_t=-\kappa y_t+by_t^3$, by expanding all functions in powers of the small parameter $b$. Results show that the symmetry is already lost at the first order in $b$ and $t$.

\subsection{Auxiliary processes} 

\subsubsection{Biased forces}

Once the right eigenfunction of the tilted generator associated with the dominant eigenvalue  is computed, e.g. $r_{\imath^v,\lambda}(v,y)\propto e^{-(1/2)[A_{\imath^v}(\lambda)v^2+B_{\imath^v}(\lambda)y^2+2C_{\imath^v}(\lambda)yv]}$ in the case of $I^v$, we can  determine the auxiliary (or driven) process that realizes  the conditioned process ${\cal P}[{\bf v}_0^t,{\bf y}_0^t\vert I^v=\imath^v t]$ in the limit $t \to\infty$. This process is a diffusion with the same two-dimensional matrix ${\bf D}'$ as the original process and a modified drift given by (cf. eq.\ (19) in [26] with $k\to -\lambda$)
\begin{align} 
\label{EqFdriven}
{\bf F}_{\imath^v,\lambda}(v,y)={\bf F}(v,y)-{\bf D}'[\lambda {\bf g}_{\imath^v}(v,y) -\nabla \ln r_{\imath^v,\lambda}(v,y)]\ .
\end{align} 
Namely, the auxiliary dynamics is governed by the coupled linear equations [eqs. (30) in the main text]
\begin{align}
\label{Eqeff} 
m\dot v_t&=-\gamma_{\rm eff} v_t -\kappa_{\rm eff}y_t+\xi_t\nonumber\\
\tau \dot y_t&=k_1 y_t+k_2 v_t+\eta_t\ .
\end{align}
with 
\begin{align} 
\label{EqFdriveniv}
\frac{\gamma_{\rm eff,\imath^v}}{\gamma} &=1+\frac{2\lambda T}{m}(\alpha_{11}-c_{11}^{-1})+\frac{2T}{m}A_{\imath^v}(\lambda)\ \ ,\ \frac{\kappa_{\rm eff,\imath^v}}{\kappa}= 1+\frac{2T}{m}\frac{\gamma}{\kappa}[\lambda\alpha_{12}+C_{\imath^v}(\lambda)] \nonumber\\
k_{1,\imath^v}&=-1-\frac{\Delta}{\tau}B_{\imath^v}(\lambda)\ \ , \ k_{2,\imath^v}=1-\frac{\Delta}{\tau}C_{\imath^v}(\lambda)\ .
\end{align}
Likewise, the coefficients of the effective drifts  for $\Sigma^m$ and $\Sigma$ are obtained as
\begin{align} 
\label{EqFdrivensigmam}
\frac{\gamma_{\rm eff,\sigma^m}}{\gamma} &=1-2\lambda+\frac{2T}{m}A_{\sigma^m}(\lambda)\ \ ,\ \frac{\kappa_{\rm eff,\sigma^m}}{\kappa}=1+\frac{2 T}{m}\frac{\gamma}{\kappa}C_{\sigma^m}(\lambda)\nonumber\\
k_{1,\sigma^m}&=-1-\frac{\Delta}{\tau}B_{\sigma^m}(\lambda) \ \ , \ k_{2,\sigma^m}=1-\frac{\Delta}{\tau}C_{\sigma^m}(\lambda)\ ,
\end{align}
and 
\begin{align} 
\label{EqFdrivensigma}
\frac{\gamma_{\rm eff,\sigma}}{\gamma} &=1+\frac{2\lambda T}{m}(\alpha_{11}-\frac{m}{T})+\frac{2T}{m}A_{\sigma}(\lambda)\ \ , \  \frac{\kappa_{\rm eff,\sigma}}{\kappa}=1+\frac{2 T}{m}\frac{\gamma}{\kappa} [\lambda \alpha_{12}+C_{\sigma}(\lambda))]\nonumber\\
k_{1,\sigma}&=-1-\frac{\Delta}{\tau}B_{\sigma}(\lambda) \ \ , \ k_{2,\sigma}=1-\frac{\Delta}{\tau}C_{\sigma}(\lambda)\ ,
\end{align}
respectively. 

\subsubsection{Biased noises}

As briefly explained in the main text, thanks to the linearity of the model, we can also define  atypical Gaussian noises $\xi_{\rm atyp}$ and $\eta_{\rm atyp}$ that create rare fluctuations of $\sigma^m$ or $\imath^v$. In the frequency domain, a solution of eqs. (\ref{Eqeff}) reads
\begin{align}
\label{EqLeff1} 
v_{\rm atyp}(\omega)&=\chi_{\rm eff}(\omega)[\xi(\omega)+\frac{\kappa_{\rm eff}}{k_1+i\omega \tau}\eta(\omega)]\nonumber\\
y_{\rm atyp}(\omega)&=-\frac{\chi_{\rm eff}(\omega)}{k_1+i\omega \tau}[k_2\xi(\omega)+(\gamma_{\rm eff}-im\omega)\eta(\omega)]
\end{align}
where $\chi_{\rm eff}(\omega)$ is given by eq.\ (32) in the main text.
By inserting this solution into the original equations of motion, we then express the two biased noises in terms of the original Gaussian white noises (eqs. (31) in the main text).
Their power spectral densities are  given by
\begin{align}
S_{\xi_{\rm atyp}}(\omega)\equiv\langle\xi_{\rm atyp}(\omega)\xi_{\rm atyp}(-\omega)\rangle&=\frac{\vert \chi_{\rm eff}(\omega)\vert^2}{k_1^2+\omega^2 \tau^2}\Big[2\gamma T [ (\gamma k_1-\kappa k_2+m\omega^2\tau)^2+\omega^2(\gamma \tau-mk_1)^2]\nonumber\\
& +\Delta[(\gamma \kappa_{\rm eff}-\gamma_{\rm eff}\kappa)^2+m^2\omega^2(\kappa-\kappa_{\rm eff})^2]\Big]\nonumber\\
S_{\eta_{\rm atyp}}(\omega)\equiv\langle\eta_{\rm atyp}(\omega)\eta_{\rm atyp}(-\omega)\rangle&=\frac{\vert \chi_{\rm eff}(\omega)\vert^2}{k_1^2+\omega^2 \tau^2}\Big[2\gamma T [ (k_1+k_2)^2+\omega^2\tau^2(1-k_2)^2]\nonumber\\
&+\Delta[(\gamma_{\rm eff}+\kappa_{\rm eff}-m\omega^2\tau)^2+\omega^2(m+\gamma_{\rm eff}\tau)^2]\Big]\ .
\end{align}
The variations of the corresponding  intensities $S_{\xi_{\rm atyp}}(\omega=0)$ and $S_{\xi_{\rm atyp}}(\omega=0)$  as a function of $\sigma^m$ and $\imath^v$ are shown in Fig. 7 of the main text.

\subsubsection{Relationship with the modified dynamics and the IFTs}\label{sec:modDynIFT}

Finally, we uncover the relationship between the auxiliary (or driven) dynamics and the modified dynamics introduced above in section A to derive the IFTs. As an example, we consider the so-called ``star" dynamics   associated with the IFT $\langle e^{-\Sigma}\rangle=1$ and defined by eq.\ (\ref{Eqauxdef}). Comparing eq.\ (\ref{EqDFTsigma}) with the equation defining the exponentially tilted path ensemble for $\Sigma$
\begin{align} 
\label{Eqtildedpath1}
{\cal P}_{\sigma,\lambda}[{\bf v}_0^t,{\bf y}_0^t]\equiv \frac{e^{-\lambda \Sigma}{\cal P}[{\bf v}_0^t,{\bf y}_0^t] }{\langle e^{-\lambda \Sigma}\rangle}\ ,
\end{align}
we readily see that 
\begin{align} 
{\cal P}^*[{\bf v}_0^t, {\bf y}_0^t]= {\cal P}_{\sigma,\lambda=1}[\tilde {\bf v}_0^t, \tilde {\bf y}_0^t] 
\end{align}
for any trajectory of duration $t$. Since the path measure of the auxiliary process becomes  equivalent to the tilted path measure as $t \to \infty$, we thus conclude that the probability to observe a trajectory with the star process and the probability to observe the time-reversed trajectory with the auxiliary process are asymptotically identical when $\lambda=1$. In particular, the corresponding stationary pdfs are simply related by time reversal
\begin{align} 
\label{Eqasympst}
p^*(v,y)=p_{\sigma,\lambda=1}(-v,y)\ .
\end{align}
 In other words, the star process and the time reversal of the auxiliary process for $\lambda=1$ must be governed by the same equations of motion in the stationary limit.  To check this identity explicitly, we  build the time reversal of the auxiliary process, carefully taking into account the fact that  $v_t$ is an odd variable. Since the  process is a two-dimensional Ornstein-Uhlenbeck process, its  time reversal is again a diffusion  governed  by the coupled equations
\begin{align} 
\label{EqFtilde}
\dot {\bf X}_t= -{\bf F}_{\sigma,\lambda}( {\bf X}_t)-{\bf D}'{\bf C}_{\sigma,\lambda}^{-1} {\bf X}_t+{\boldsymbol \xi}_t\ ,
\end{align} 
where ${\bf X}_t=(-v_t,y_t)$, ${\bf F}_{\sigma,\lambda}({\bf X}_t)$ is the two-dimensional force corresponding to the effective drifts defined  by eqs. (\ref{EqFdrivensigma}), and ${\bf C}_{\sigma,\lambda}$ is the stationary covariance matrix in the auxiliary process, hence $p_{\sigma,\lambda}(v,y)=l_{\sigma,\lambda}(v,y)r_{\sigma,\lambda}(v,y)\propto  e^{-(1/2){\bf X}^T{\bf C}_{\sigma,\lambda}^{-1}{\bf X}}$. We then set $\lambda=1$ in these equations and use the fact that $\l_{\sigma,\lambda=1}(v,y)=p(v,y)$ (see the remark at the end of section C.1.a).  Therefore,
\begin{align} 
\label{Eqpdfdriven1}
p_{\sigma,\lambda=1}(v,y)\propto e^{-\frac{1}{2}\big\{[\alpha_{11}+A_{\sigma}(1)]v^2+[\alpha_{22}+B_{\sigma}(1)]y^2+2[\alpha_{12}+C_{\sigma}(1)]yv\big\}}\ , 
\end{align}
which gives  the expression of ${\bf C}_{\sigma,\lambda=1}^{-1}$. Inserting  into eqs. (\ref{EqFtilde}), we find that the terms involving the quantities $A_{\sigma}(1),B_{\sigma}(1)$ and $C_{\sigma}(1)$ cancel out and we finally recover the equations of motion of the star process.

We stress that the asymptotic equivalence between the two processes is only valid when the right and left eigenfunctions $r_{\sigma,\lambda=1}(v,y)$ and $l_{\sigma,\lambda=1}(v,y)$  are normalizable and the pre-exponential factor $g_{\sigma}(\lambda=1)$ is finite. This latter condition  is not satisfied when the stationary state of the star process does not exist (i.e. $\int dv\: dy\: p^*(v,y)$ diverges). Then  eq.\ (\ref{Eqasympst}) does not hold and $\l_{\sigma,\lambda=1}(v,y)\ne p(v,y)$. However, it is noteworthy that the stationary pdf of the auxiliary process $p_{\sigma,\lambda=1}(v,y)= l_{\sigma,\lambda=1}(v,y)r_{\sigma,\lambda=1}(v,y)$ is still normalizable.


\begin{thebibliography}{10}
\bibitem{note2} {Note that the initial state distribution of the backward path is $p_t(v_t,y_t)$, the final state distribution of the forward path, and not  $p_t(-v_t,y_t)$.}
\bibitem{note1} On the other hand, the symmetry is preserved if one adds a force {depending linearly on the position $x_t$ of the Brownian particle in the first Langevin equation.}
\end{thebibliography}
\end{document}